\documentclass[11pt,preprint]{aastex}
\usepackage{graphics}
  
\begin{document}

\let\hat\widehat

\title{An Unbiased Method of Modeling the Local Peculiar Velocity Field with Type Ia Supernovae}
\author{
{Anja Weyant}\altaffilmark{1}, 
{Michael Wood-Vasey}\altaffilmark{1}, 
{Larry Wasserman}\altaffilmark{2}, 
{Peter Freeman}\altaffilmark{2}
}

\altaffiltext{1}{Department of Physics and Astronomy, University of Pittsburgh, Pittsburgh, PA 15260}
\altaffiltext{2}{Department of Statistics, Carnegie Mellon University, Pittsburgh, PA 15213}

\email{anw19@pitt.edu, wmwv@pitt.edu, larry@stat.cmu.edu, pfreeman@cmu.edu}

\begin{abstract}
We apply statistically rigorous methods of nonparametric risk estimation to the problem of inferring the local peculiar velocity field from nearby supernovae (SNIa).  We use two nonparametric methods - Weighted Least Squares (WLS) and Coefficient Unbiased (CU) - both of which employ spherical harmonics to model the field and use the estimated risk to determine at which multipole to truncate the series.  We show that if the data are not drawn from a uniform distribution or if there is power beyond the maximum multipole in the regression, a bias is introduced on the coefficients using WLS.  CU estimates the coefficients without this bias by including the sampling density making the coefficients more accurate but not necessarily modeling the velocity field more accurately. After applying nonparametric risk estimation to SNIa data, we find that there are not enough data at this time to measure power beyond the dipole. The WLS Local Group bulk flow is moving at $538 \pm 86$~km~s$^{-1}$ towards ($l,b$) $=$ ($258^{\circ} \pm 10^{\circ}, 36^{\circ} \pm 11^{\circ}$) and the CU bulk flow is moving at $446 \pm 101$~km~s$^{-1}$ towards ($l,b$) $=$ ($273^{\circ} \pm 11^{\circ}, 46^{\circ} \pm 8^{\circ}$).  
We find that the magnitude and direction of these measurements are in agreement with each other and previous results in the literature.
\end{abstract}


\section{Introduction}

Inhomogeneities in the matter distribution of our universe perturb the motions of individual galaxies from the overall smooth Hubble expansion.  These motions, called peculiar velocities, result from gravitational interactions with the spectrum of fluctuations in the matter-density and are therefore a direct probe of the distribution of dark matter.  The peculiar velocity of an object is influenced by matter on all scales and modeling the peculiar velocity field allows one to probe scales larger than the sample.   Calculating the dipole moment of the peculiar velocity field, or bulk flow, is an example of a measurement which helps us investigate the density fluctuations on large scales.  Fluctuations in density on many Mpc scales are well described by linear physics and can be used to probe the mass power spectrum while fluctuations on small scales become highly non-linear and difficult to model. 

From an accurate measurement of the local peculiar velocity field we can infer the properties of the dark matter distribution.  On scales $\gtrsim 10$~Mpc we can use linear perturbation theory to estimate the bias free mass power spectrum directly from 
\begin{equation} U({\bf r }) = \frac{H_0 \Omega_m^{0.6}}{4 \pi} \int d^3{\bf r^{\prime}} \frac{\delta_m({\bf r^{\prime}})({\bf r^{\prime}}-{\bf r})}{\mid {\bf r^{\prime}}-{\bf r} \mid^3}\end{equation}
where $\delta_m({\bf r})$ is the density contrast defined by $(\rho-\bar {\rho})/(\bar {\rho})$, $\bar \rho$ is the average density, and $\Omega_m$ is the matter density parameter \citep{peebles93}.  In the past, measurements of the matter power spectrum using 
galaxy peculiar velocity catalogs consistently produced power spectra with large amplitudes \citep{Zaroubi97, Freudling99, Zaroubi01}.  A renewed interest in bulk flow measurements has recently produced power spectra with lower amplitudes, which is often characterized by $\sigma_8$, which are consistent with WMAP \citep{Park06, Feldman08, Abate09, Song10, lavaux10} and some which challenge the $\Lambda$CDM cosmology \citep{Kashlinsky08, Watkins09, Feldman10, Macaulay10}.

Peculiar velocity measurements also enable one to measure the matter distribution independent of redshift surveys.  This allows for comparison between the galaxy power spectrum and matter power spectrum to probe the bias parameter $\beta$ which specifies how galaxies follow the total underlying matter distribution \citep{Pike05, Park06, Davis10}.  In addition to measuring $\beta$, this comparison can also be used to test the validity of the treatment of bias as a linear scaling \citep{Abate08}.

By accurately measuring the local velocity field it is also possible to limit its effects on derived cosmological parameters \citep{Cooray06,HuiGreene06,Gordon07,Neill07,DavisT10}.
The basic cosmological utility of Type Ia Supernovae (SNIa) comes from comparing an inferred luminosity distance with a measured redshift.  This redshift is assumed to be from cosmic expansion.  However, on smaller physical scales where large scale structure induces peculiar velocities that create large fluctuations in redshift, this measured redshift becomes a combination of cosmic expansion and local bulk motion and thus of limited utility in inferring cosmological parameters from the corresponding luminosity distance. 
While traditionally this troublesome regime has been viewed to extend out to $z<0.05$, recent work has shown significant effects out to $z < 0.1$ \citep{Cooray06,HuiGreene06}.  Hence peculiar velocities from SNIa add scatter to the Hubble diagram in the nearby redshift regime which adds uncertainty to derived cosmological parameters, including the dark energy equation-of-state parameter.  In an ongoing effort to probe the nature of dark energy, surveys such as the CfA Supernova Group\footnote{\url{http://www.cfa.harvard.edu/supernova/SNgroup.html}}\citep{Hicken09a}, SNLS\footnote{\url{http://cfht.hawaii.edu/SNLS/}} \citep{astier06}, Pan-STARRS\footnote{\url{http://pan-starrs.ifa.hawaii.edu/public/}}, ESSENCE\footnote{\url{http://www.ctio.noao.edu/essence/}}~\citep{miknaitis07}, Carnegie Supernova Project (CSP) \footnote{\url{http://csp1.lco.cl/~cspuser1/PUB/CSP.html}}~\citep{hamuy06,Folatelli10}, the Lick Observatory Supernova Search KAIT/LOSS~\citep{filippenko01,Leaman10}, Nearby SN Factory\footnote{\url{http://snfactory.lbl.gov/}}~\citep{aldering02}, SkyMapper\footnote{\url{http://www.mso.anu.edu.au/skymapper/}}~\citep{Murphy09}, Palomar Transient Factory\footnote{\url{http://www.astro.caltech.edu/ptf/}}~\citep{law09} hope to obtain tighter constraints on cosmological parameters. With the future influx of SNIa data, statistical errors will be reduced but an understanding of systematic errors is required to make improvements on cosmological measurements \citep{DETF}.  Averaging over many SNIa reduces scatter caused by random motions but not those caused by coherent large scale motions.  One expects these bulk motions to converge to zero with increasing volume in the rest frame of the CMB with the rate of convergence depending on the amplitude of the matter perturbations.  This fact motivates determining both the monopole and dipole component of the local peculiar velocity field \citep{Zaroubi02}.  

To model the local peculiar velocity field or flow field requires a measure of an object's peculiar velocity and its position on the sky.  The peculiar velocity of an object, such as a galaxy or Type Ia Supernova (SNIa), given the redshift $z$ and cosmological distance $d$ is
\begin{equation} U=H_0d_l(z)-H_0d\end{equation}
where $H_0$ is the Hubble parameter and $H_0d_l(z)$ is the recessional velocity described by
\begin{equation} H_{0}\,d_{l}(z) = c (1+z) \int_{0}^{z}\left[ \Omega_M(1+{z'})^3+\Omega_{\Lambda} \right]^{-1/2}d{z'}.\end{equation}
Redshifts to galaxies can be measured accurately with an error $\sigma_z\sim0.001$. Therefore, the accuracy of a measure of an object's peculiar velocity rests on how well we can determine its distance.  

A variety of techniques exist to determine $d$.  Distances to spiral galaxies can be measured through the Tully-Fisher (TF) relation \citep{Tully77} which finds a power law relationship between the luminosity and rotational velocity.  This method has been one of the most successful in generating large peculiar velocity catalogs.  The SFI++ dataset \citep{Masters06, Springob07} for example, is one of the largest homogeneously derived peculiar velocity catalog using I-band TF distances to $\sim$~5000 galaxies with $\sim 15$\% distance errors.  This catalog builds on the Spiral Field I-band (SFI; \citet{Giovanelli94, Giovanelli95, daCosta96, Haynes99a, Haynes99b}), Spiral Cluster I-band (SCI; \citet{Giovanelli97a, Giovanelli97b}) and the Spiral Cluster I-band 2 (SC2; \citet{Dale99a, Dale99b}) catalogs.  The Two Micron All-Sky Survey (2MASS) Tully-Fisher (2MTF) survey \citep{Masters08} aims to measure TF distances to all bright spirals in 2MASS in the J, H, and K bands.  The Kinematics of the Local Universe catalog (KLUN)\footnote{\url{http://klun.obs-nancay.fr/}}  \citep{Theureau98}, which is the only catalog to exceed SFI++ in number of galaxies, consists of B-band TF distances to 6600 galaxies.  The velocity widths in this catalog are not homogeneously collected and measured adding to the errors.  Additionally, the B-band TF relationship exhibits more scatter -- which translates to larger distance errors -- than the I, J, H, and K bands due to Galactic and internal extinction, making the SFI++ arguably the best galaxy peculiar velocity catalog currently available.  Finally, with the Square Kilometer Array (SKA) \citep{Dewdney09} we expect TF catalogs to grow out to larger distances using less HI in the near future.  Using TF derived peculiar velocities, measurements of the bulk flow and shear moments have been made \citep{Giovanelli98, Dale99c, Courteau00, Kudrya03,Feldman08, Kudrya09, Nusser11} and cosmological parameters have been constrained \citep{daCosta98, Freudling99, Borgani00, Branchini01, Pike05,Masters06, Park06,  Abate09, Davis10}.

Fundamental Plane (FP) distances \citep{Djorgovski87} which express the luminosity of an elliptical galaxy as a power law function of its radius and velocity dispersion also enable one to generate large peculiar velocity catalogs.  Several smaller datasets utilize this distance indicator. The Streaming Motions of Abell Clusters (SMAC) project \citep{Smith00,Hudson01, Smith01} is an all-sky FP survey of 699 galaxies with $\sim 20$\% distance errors.  The EFAR project \citep{Wegner96, Colless01} studied 736 elliptical galaxies in clusters in two regions on the sky to improve distance estimates to 85 clusters.   Larger FP programs include the NOAO Fundamental Plane Survey \citep{Smith04} which will provide FP measurements to $\sim$4000 early-type galaxies and the 6 Degree Field Galaxy Survey (6dFGS) \citep{Wakamatsu03, Jones09}; a southern sky survey which, in combination with 2MASS, hopes to deliver more than 10,000 peculiar velocities using near infrared FP distances.  The increase in the number of objects makes this catalog competitive even though individual distance errors are greater than TF errors.  The $D_n-\sigma$ relation is a reduced parameter version of FP with typical errors of $\sim 25$\%\citep{Dressler87}.  The Early-type NEARby galaxies (ENEAR) \citep{daCosta00a, Bernardi02} project measured galaxy distances based on $D_N-\sigma$ and FP to 1359 and 1107 galaxies and the Mark III Catalog of Galaxy Peculiar Velocities \citep{Willick95, Willick96, Willick97} contains 3300 galaxies with estimated distances from TF and $D_n-\sigma$.  Global features of large-scale motions \citep{Dressler87b, Hudson99, Dekel99, daCosta00b,Hudson04, Colless01} and derived parameters have been measured using these catalogs \citep{Davis96, Park00, Rauzy00, Zaroubi01, Nusser01, Hoffman01}.

Other distance measurements to galaxies are more difficult to make or not as precise.  \citet{Tonry01} report distances to 300 early type galaxies using Surface Brightness Fluctuations (SBF) whose observations were obtained over a period of $\sim 10$ years.  This method measures the luminosity fluctuations in each pixel of a high signal-to-noise CCD image of a galaxy where the amplitude of these fluctuations is inversely proportional to the distance. A $\sim 5$\% distance uncertainty can be obtained under the best observing conditions \citep{Tonry00} making SBF a useful method for $cz<4000$~km/s.  Although it is difficult to create a large catalog of objects, \citet{Blakeslee99} used SBF distances to put constraints on $H_0$ and $\beta$.  Tip of the Red Giant Branch (TRGB) \citep{Madore95} and Cepheid distances are challenging to obtain as one must have resolved stars which limits observations to the local Universe.

SNIa are ideal candidates to measure peculiar velocities because they have a standardizable brightness and thus accurate distances can be calculated with less than 7\% uncertainty \citep[e.g.][]{JRK07}.   Only recently through the efforts like the CfA Supernova Group, LOSS, and CSP have there been enough nearby SNIa ($\sim 400$) to make measurements of bulk flows.
Measurements of the monopole, dipole, and quadrupole have been made which find dipole results compatible with the CMB dipole \citep{Colin10, Haugbolle07, JRK07, Giovanelli98, Riess95}.  Measurements of the monopole as a function of redshift have been used to test for a local void \citep{zehavi98,JRK07}.  SNIa peculiar velocity measurements have also been used to put constraints on power spectrum parameters \citep{Radburn04, Watkins07}. \citet{Hannestad08} forecast the precision with which we will be able to probe $\sigma_8$ with future surveys like LSST.   Following up on \citet{Cooray06,HuiGreene06}, \citet{Gordon08} and \citet{DavisT10} investigated the effects of correlated errors when neighboring SNIa peculiar velocities are caused by the same variations in the density field.  Not accounting for these correlations underestimates the uncertainty as each new SNIa measurement is not independent.  Recent investigations in using near-infrared measurements of SNIa to measure distances have shown promise for better standard candle behavior and the potential for more accurate and precise distances to galaxies in the local Universe \citep{Krisciunas04, WV08, Mandel09}.

A wide range of methods have been developed to model the local peculiar velocity field.  \citet{Nusser95} present a method for deriving a smooth estimate of the peculiar velocity field by minimizing the scatter of a linear inverse Tully-Fisher relation $\eta$ where the magnitude of each galaxy is corrected by a peculiar velocity.  The peculiar velocity field is modeled in terms of a set of orthogonal functions and the model parameters are then found by maximizing the likelihood function for measuring a set of observed $\eta$.  This method was applied to the Mark III \citep{Davis96} and SFI \citep{daCosta98} catalogs.
Several other methods have been developed and tested on e.g., SFI and Mark III catalogs to estimate the mass power spectrum and compare peculiar velocities to galaxy redshift surveys which utilize or compliment rigorous maximum likelihood techniques \citep{Willick98, Freudling99, Zaroubi99, Hoffman00, Zaroubi01, Branchini01}.  Nonparametric models \citep{Branchini99} and orthogonal functions \citep{Nusser94, Fisher95} have been implemented to recover the velocity field from galaxies in redshift space.  Smoothing methods  \citep{Dekel99, Hoffman01} have also been used to tackle large random errors and systematic errors associated with nonuniform and sparse sampling.  More recently, \citet{Watkins09} introduce a method for calculating bulk flow moments which are comparable between surveys by weighting the velocities to give an estimate of the bulk flow of an idealized survey.  The variance of the difference between the estimate and the actual flow is minimized.  \citet{Nusser11} present ACE (All Space Constrained Estimate), a three dimensional peculiar velocity field constrained to match TF measurements which is used to reconstruct the bulk flow.


Comparisons between different SNIa and galaxy surveys and methods show that measurements of the local velocity field are highly correlated and in agreement \citep{Zaroubi02,Hudson03, Hudson04,Radburn04, Pike05,Sarkar07,Watkins07}.  However, peculiar velocity datasets which have recently been combined (namely \citet{Feldman10}, but \citet{Watkins09} and \citet{Ma10} also combine datasets) are in disagreement with \citet{Nusser11}.  Errors in distance measurements and the non-uniform sampling of objects across the sky due to the Galactic disk ($\sim$40\% of the sky) aggravate the systematic errors \citep{Zaroubi02}. These systematic errors cause inconsistencies among the different models and must be dealt with in a statistically sound fashion.  Since the field is at a point where rough agreement exists between the different methods, and modeling can be done with better precision as the amount of data continues to increase, it is time to investigate the systematic errors that limit our measurements.

In this paper we present a statistical framework that can be used to properly extract the available flow field from observations of nearby SNIa, while avoiding the historical pitfalls of incomplete sampling and over-interpretation of the data.  In particular, we emphasize the distinction between finding a best overall model fit to the data and finding the best unbiased value of a particular coefficient or set of coefficients of the model, e.g., the direction and strength of a dipole term due to our local motion.  The first task is to provide a framework for modeling the peculiar velocity field which adequately accounts for sampling bias due to survey sky coverage, galactic foregrounds, etc.  These methods are discussed in \S\ref{sec:NP}.  We then introduce risk estimation as a means of determining where to truncate a series of basis functions when modeling the local peculiar velocity field, e.g., should we fit a function out to a quadrupole term.  Risk estimation is a way of evaluating the quality of an estimate of the peculiar velocity field as a function of $l$ moment, whose minimum determines the optimal balance of the bias and variance.   These methods are outlined in \S\ref{sec:risk}.   In \S\ref{sec:sim} and \S\ref{sec:data} we apply these methods to a simulated dataset and SNIa data pulled from recent literature and discuss our results.  We then apply our methods to simulated data modeled after the actual data and examine their performance as we alter the direction of the dipole in \S\ref{sec:dipoleAnalysis}.  In \S\ref{sec:conclusion} we conclude and present suggestions for future work.

\section{Nonparametric Analysis of a Scalar Field}\label{sec:NP}

A peculiar velocity field at a given redshift can be written as 
 \begin{equation}U_n=f(x_n)+\epsilon_n\end{equation}  
where $U_n$ is the observed peculiar velocity at position
$x_n=(\theta_n,\phi_n)$ on the sky, $\epsilon_n$ is the observation error,
and $f$ is our peculiar velocity field.  Since we expect $f$ to be a
smoothly varying function across the sky, it can be decomposed as
\begin{equation} f(x) =\displaystyle\sum_{j=0}^{\infty}{\beta_j\phi_j(x)}
\end{equation}
where $\phi_j$, [$j=0,1,2...$] forms an orthonormal basis and $\beta_j$ is
given by \begin{equation}\beta_j=\int{\phi_j(x)f(x)dx}.\end{equation}

In this work we apply the real spherical harmonic basis as we are physically interested in a measurement of the dipole and follow a
procedure similar to \citet{Haugbolle07}. The radial velocity field on a
spherical shell of a given redshift can be expanded using spherical
harmonics:
\begin{eqnarray}
f&=&\displaystyle\sum_{l=0}^{\infty}\displaystyle\sum_{m=-l}^{l}a_{lm}Y_{lm} \\
  &=&\sum_{l=0}^{\infty} \left\{ \sum_{m=1}^{l}(a_{l,-m}Y_{l,-m}+a_{lm}Y_{lm})+a_{l0}Y_{l0} \right\}. 
\end{eqnarray}
Using $a_{l,-m}=(-1)^ma_{lm}^*$ and $Y_{l,-m}=(-1)^mY_{lm}^*$ the
expansion for the real radial velocity can be rewritten as
\begin{eqnarray}
f &=&\displaystyle\sum_{l=0}^\infty \left\{ \displaystyle\sum_{m=1}^l\left[2\Re(a_{lm}Y_{lm})\right]+a_{l0}Y_{l0} \right\}\\
&=&\displaystyle\sum_{l=0}^{\infty} \left\{ \displaystyle\sum_{m=1}^l\left[2\Re(a_{lm})\Re(Y_{lm})-2\Im(a_{lm})\Im(Y_{lm})\right]+a_{l0}Y_{l0} \right \}.
\end{eqnarray}
Our real orthonormal basis is then
$[Y_{l0},{\sqrt{2}\Re(Y_{lm}),-\sqrt{2}\Im(Y_{lm})},m=1,...,l]$.

We have peculiar velocity measurements for a
finite number of positions on the sky and therefore cannot fit an infinite set
of smooth functions. We estimate $f$\footnote{Following statistical practice we denote an estimated quantity with a hat.} by
\begin{equation}\label{fhat} \hat f(x)=\displaystyle\sum_{j=0}^J{\beta_j\phi_j(x)}\end{equation}
where $J$ is a tuning parameter, more precisely the $l^{\rm th}$ moment that
we fit out to.  By introducing a tuning parameter our methods are nonparametric; we do not a priori decide where to truncate the series of spherical harmonics but allow the data to determine the tuning parameter.  Our task is now to estimate $\beta$ and determine $J$.

\subsection{Weighted Least Squares Estimator $\hat \beta_J$}

Consider an ideal case where the 2D peculiar velocity field can be represented {\it exactly} as a finite sum of spherical harmonics, plenty of uniformly distributed data is sampled, and the true $J$ is chosen as a result.  The Gauss-Markov theorem (see, e.g., \citealt{Hastie09}) tells us that the best linear unbiased estimator with minimum variance for a linear model in which the errors have expectation zero and are uncorrelated is the weighted least squares (WLS) estimator. 
If we define $Y_J$ as the $N \times J$ matrix
\begin{equation}
Y_J=
\left[ {\begin{array}{cccc}
\phi_0(x_1) & \phi_1(x_1) & \cdots&\phi_J(x_1)\\
\phi_0(x_2)&\phi_1(x_2)&\cdots&\phi_J(x_2)\\
\vdots&\vdots &\cdots &\cdots\\
\phi_0(x_N)&\phi_1(x_N)&\cdots &\phi_J(x_N)\\
\end{array} } \right]
\end{equation}
and the column vectors $\beta_J = (\beta_0,...,\beta_J)$, $U=(U_1,...,U_N)$ and $\epsilon =
(\epsilon_1,...,\epsilon_N)$  we can then write 
\begin{equation}U = Y_J \beta_J + \epsilon .\end{equation}
The WLS estimator, $\hat{\beta_J}$, that minimizes the residual sums of squares is  
\begin{equation}\label{betahat} \hat\beta_J = (Y_J^TWY_J)^{-1}Y_J^TWU \end{equation}
where the diagonal elements of $W$ are equal to one over the variance and the off-diagonal elements are zero. 

Any estimate for $\hat f$ which truncates an infinite series of functions will produce the same overall bias on the peculiar velocity field, namely 
\begin{equation}f-\langle \hat f \rangle  = \displaystyle\sum_{j=J+1}^{\infty}{\beta_j\phi_j}\end{equation}
where ``$\langle$ $\rangle$'' denotes the ensemble expectation value. However, in the case we are considering there is no power at multipoles beyond $j=J$ so this bias will go to zero.  

The estimates of the coefficients $\hat \beta_J$ are also unbiased if the correct tuning parameter is chosen.  If $Y_{\infty}$ and $\beta_{\infty}$ are defined over the range $[J+1, \infty)$ then the bias on $\hat \beta_J$ (Appendix~\ref{appendix:biasWLS})
\begin{equation}
\beta_J-\langle \hat\beta_J \rangle = \left\langle (Y_J^TWY_J)^{-1}Y_J^TW(Y_{\infty}\beta_{\infty}) \right\rangle
\end{equation}
is a function of all the $\beta$'s beyond the tuning parameter, i.e., the lower-order modes are contaminated by power at higher $l$'s.  For this case there is no power at multipoles beyond $j=J$, $\beta_{\infty}=0$ and our coefficients are unbiased. 

\subsection{Coefficient Unbiased Estimator $\hat \beta_j^*$}

If our data are not well-sampled, i.e., not drawn from a uniform distribution, or if spherical harmonics are not a good representation of the true velocity field then it is possible that there will be power beyond the best tuning parameter.  This does not indicate a failure in determining the tuning parameter via risk (see \S\ref{sec:risk}) but is a consequence of the data.  

Ideally we would like to obtain the unbiased coefficients because we tie physical meaning to the monopole and dipole.   Suppose our dataset $x$ is sampled according to a sampling density $h(x)$ which quantifies how likely one is to sample a point at a given position on the sky, then 
\begin{eqnarray}\label{eq:CU}
\left \langle \frac{U\phi_j(x)}{h(x)} \right \rangle &=& \left \langle \frac{(f(x)+\epsilon)\phi_j(x)}{h(x)} \right \rangle \\
&=&\left \langle \frac{f(x)\phi_j(x)}{h(x)} \right \rangle\\
&=&\int{\frac{f(x)\phi_j(x)}{h(x)}h(x)dx}\\
&=&\beta_j.
\end{eqnarray}
A weighted unbiased estimate of $\beta_j$ is therefore (see Appendix~\ref{appendix:biasCU})
\begin{equation}\label{CUbeta} \hat\beta_j^* =
\frac{\displaystyle\sum_{n=1}^N{\frac{U_n\phi_j(x_n)}{h(x_n)\sigma_n^2}}}{\displaystyle\sum_{n=1}^N \frac{1}{\sigma_n^2}}\end{equation}
where $\sigma_n$ is the uncertainty on the peculiar velocity.  We will call this our coefficient-unbiased (CU) estimate, $\hat \beta_j^*$.  

Although the CU estimate is unbiased, its accuracy depends on the sampling density.  The sampling density in most cases is unknown and must be estimated from the data.

\subsubsection{Estimating $h(x)$}\label{sec:h}

The sampling density is a normalized scalar field which can be modeled several ways.  We outline the process using orthonormal basis functions and will continue to use the real spherical harmonic basis, $\phi$.  We decompose $h$ as
\begin{equation}\label{h} h(x) = \displaystyle\sum_{i=0}^{\infty}\alpha_i\phi_i(x) \simeq \displaystyle\sum_{i=0}^{I}Z_i\phi_i(x) \end{equation}
where $I$ is the tuning parameter.  We estimate $\alpha_i$ by
\begin{equation} Z_i = \frac{1}{N}\displaystyle\sum_{n=1}^{N}\phi_i(x_n)\end{equation}
since 
\begin{equation}  \langle Z_i \rangle  = \int{\phi_i(x)h(x)dx} = \alpha_i .\end{equation}

It is difficult to create a normalized positive scalar field using a truncated set of orthogonal functions.  In practice there may be patches on the sky which have negative $\hat h$.  Since a negative sampling density has no physical meaning, we set all negative regions to zero, add a small constant offset component to the sampling density and renormalize.  This will prevent division by zero when a data point lies in a negative $\hat h$ region. This procedure adds a small bias but is a standard practice when using orthogonal functions and small datasets (see, e.g., \citealt{Efromovich}).

Is a spherical harmonic decomposition of the sampling density appropriate?  
While using an orthogonal basis is desirable, the choice of spherical harmonics to model a patchy sampling density is clearly non-ideal.
Smoothing the data with a Gaussian kernel or using a wavelet decomposition to model $h$ would be a good alternative, especially when the distribution of data is sparse or if there are large empty regions of space.  We merely present the formalism to determine $h$ with orthonormal functions and encourage astronomers to use sampling density estimation with any basis set.

\section{Determining Tuning Parameter via Risk Estimation}\label{sec:risk} 

 Recall that the tuning parameter determines at which $l$ moment to truncate the series of spherical harmonics. We determine this value by minimizing the estimated risk.  The risk is a way of evaluating the quality of a nonparametric estimator by balancing the bias and variance (Appendix~\ref{appendix:risk}) which determines the complexity of the function we fit to the data.  If the bias is large and the variance is small, the function will be too simple, under-fitting the data.  This would be analogous to only using a monopole term when there is power at higher $l$.   If the opposite is true, the data are over-fitted, similar to fitting many spherical harmonics in order to describe noisy data.   

\subsection{Risk Estimation for WLS}  

Recall the estimated peculiar velocity field
\begin{equation}\hat f=Y_J \hat \beta_J \equiv LU\end{equation}
where we introduce the smoothing matrix $L$ 
\begin{equation}L=Y_J(Y_J^TWY_J)^{-1}Y_J^TW.\end{equation}
Note that the $n \rm ^{th}$ row of the smoothing matrix is the effective kernel for estimating $f(x_n)$. 
The risk is the integrated mean squared error
\begin{equation}\label{trueR} R(J) = \left\langle \frac{1}{N} \displaystyle \sum_{n=1}^{N}(f(x_n)-\hat f(x_n))^2 \right\rangle \end{equation}
and can be estimated by the leave-one-out cross-validation score 
\begin{equation}\hat R(J) = \frac{1}{N}\displaystyle\sum_{n=1}^{N}{(U_n-\hat f_{(-n)}(x_n))^2}\end{equation}
where $\hat f_{(-n)}$ is the estimated function obtained by leaving out the $n^{th}$ data
point (see, e.g., \citealt{Wasserman}).  For a linear smoother in which $\hat f$ can be written as a linear sum of functions, the estimated risk can be written in a less computationally expensive form 
\begin{equation}\label{loocv}\hat R(J) = \frac{1}{N}\displaystyle\sum_{n=1}^{N} \left( {\frac{U_n-\hat f(x_n)}{1-L_{nn}}} \right)^2\end{equation}
where $L_{nn}$ are the diagonal elements of the smoothing matrix.     

There are a few important things to note.  First, the risk gives the best tuning parameter to use in order to model the entire function, e.g., the peculiar velocity field over the entire sky for a given set of data.  This is different than claiming the most accurate component of the field, e.g., the best measurement of the dipole.  Secondly, the accuracy to which Eq.~\ref{loocv} estimates the risk depends on the number of data points used and will be better estimated with larger datasets.  Finally, the value of the estimated risk changes for different datasets.  What is important for comparison are the relative values of the risk for different tuning parameters.  Although not explored in this paper, one can also use the estimated risk to compare bases with which one could model the peculiar velocity field. 

\subsection{Risk Estimation for CU}
We start by calculating the variance and bias on $h$.  The variance on $\hat h$ is the estimated variance on the coefficients $Z_i$ given by
\begin{equation}\label{evar}\hat \sigma_i^2 = \frac{1}{N^2}\displaystyle\sum_{n=1}^{N}(\phi_i(x_n)-Z_i)^2.\end{equation}
The bias on $h$ by definition is
\begin{equation}
h-\langle \hat h \rangle  = \displaystyle\sum_{i=I+1}^{\infty}\alpha_i\phi_i.
\end{equation}
We can only calculate the bias out to the maximum number of independent basis functions, $L$, less than the number of data points.  For spherical harmonics, this is given by 
\begin{equation} \displaystyle \sum_{l=0}^L 2l+1 \le N.\end{equation} 
The risk of the estimator is then the variance plus the bias squared
\begin{equation} \hat R(I) = \displaystyle\sum_{i=0}^I\hat\sigma_i^2+\displaystyle\sum_{i=I+1}^{L}(Z_i^2-\hat\sigma_i^2)_+\end{equation}
where we have used  Eq.~\ref{evar} to replace the bias squared $\alpha_i^2$ with $Z_i^2-\hat \sigma_i^2$ 
and $+$ denotes only the positive values.  

Estimating the risk for CU is similar to WLS using Eq.~\ref{loocv}.  $\hat f(x_i)$ must now be calculated with the unbiased coefficients $\hat\beta_j^*$ and the diagonal elements of the smoothing matrix $L_{nn}$ (see Appendix~\ref{appendix:L}) are
\begin{equation} L_{nn}=\frac{\displaystyle\sum_{j=0}^J\frac{\phi_j^2(x_n)}{\hat h(x_n)\sigma_n^2}}{\displaystyle\sum_{n=1}^N \frac{1}{\sigma_n^2}}.\end{equation}

\section{Application to Simulated Data}\label{sec:sim}

To compare WLS and CU we created a simulated dataset with a non-uniform distribution and a known 2D peculiar velocity field.  We discuss how the dataset is created followed by an application of each regression method and a discussion comparing the methods. 

\subsection{Simulated Data}

\begin{figure}[ht]
\begin{center}
\includegraphics[totalheight=2.5in]{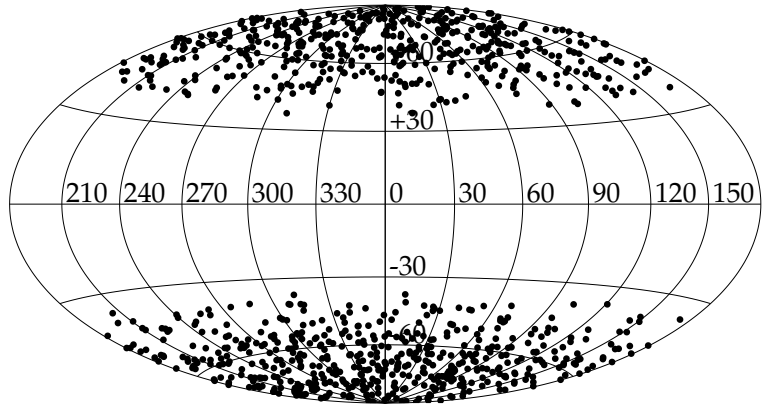}\\
\caption{2D distribution of 1000 simulated data points.  All plots are in galactic coordinates. The distribution was created from the sampling density $h=NY_{20+}$ where $N$ is a normalization factor and $+$ indicates positive values.  All negative values in the sampling density are set to zero.  Although this $h$ is not physical, the large empty galactic plane is ideal for testing the methods.}
\label{fig:H}
\end{center}
\end{figure}

We built the dataset with a non-uniform $h$ using rejection sampling.  
To do this we start with a uniform distribution of points over the entire sky and evaluate a non-uniform sampling density at each point according to 
\begin{equation}h = NY_{20+} \end{equation}
where $N$ is a normalization factor and $+$ indicates only positive values.  This has the effect of ``masking'' a region of the sky.  We then choose the 1000 most likely points given some random ``accept'' parameter.  If the accept value is less then the sampling density value, that point is selected.  A typical distribution of data points is shown in Figure~\ref{fig:H}.  This pathological sampling density provides a useful demonstration of the methods.

\begin{figure}[ht]
\begin{center}
\includegraphics[totalheight=2.5in]{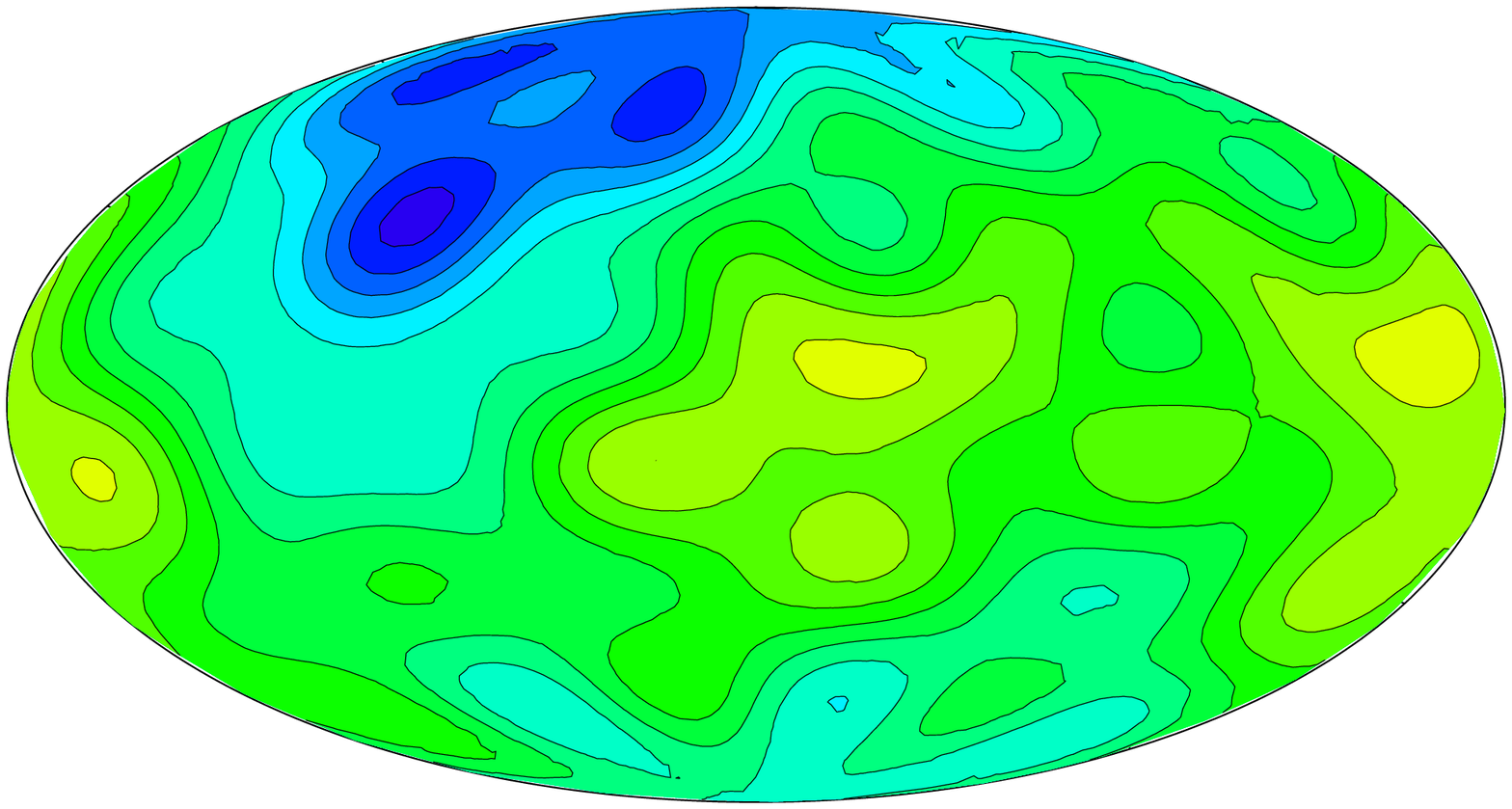}
\includegraphics[totalheight=2.5in]{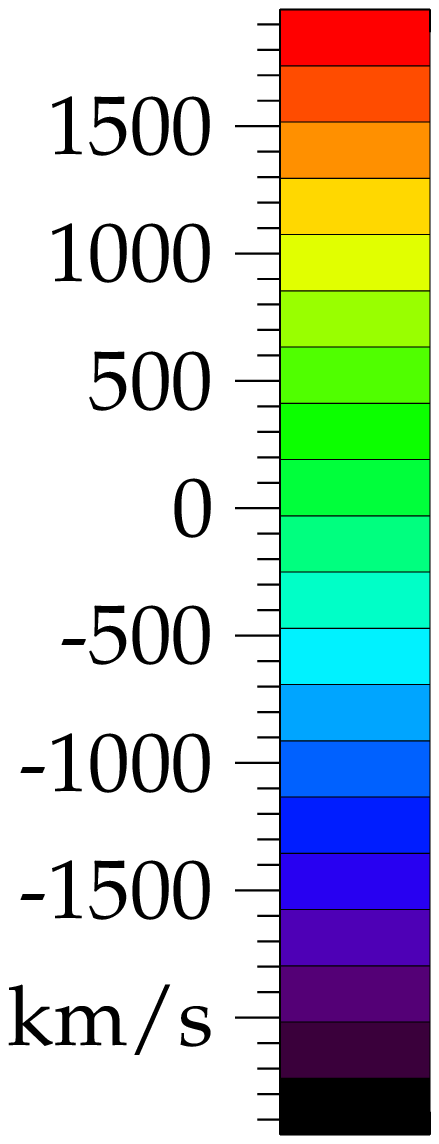}\\
\caption{2D simulated peculiar velocity field in $\rm km$ $\rm s^{-1}$, described by Eq.~\ref{vTrue}.}
\label{fig:vTrue}
\end{center}
\end{figure}

The simulated real 2D peculiar velocity field is described by  
\begin{eqnarray}\label{vTrue} V &=& 180Y_{00}-642Y_{10}-1000Y_{20}+\Re(-38Y_{11}+1061Y_{22}+150Y_{86}+300Y_{76}) \nonumber\\
&&-\Im(1146Y_{11}+849Y_{21}+707Y_{83}) \end{eqnarray}
and is shown in Figure~\ref{fig:vTrue}.  
We assign an error to each data point of 350 km $\rm s^{-1}$ and Gaussian scatter the peculiar velocity appropriately.  This error includes the error on the measurement of the magnitude, $\sigma_{\mu}$, the redshift error, $\sigma_z$, and a thermal component of $\sigma_v=300$~km~s$^{-1}$ attributed to local motions of the SNIa \citep{JRK07}. 

\subsection{Recovered Peculiar Velocity Field from WLS}

\begin{figure}[ht]
\begin{center}
\includegraphics[totalheight=3in]{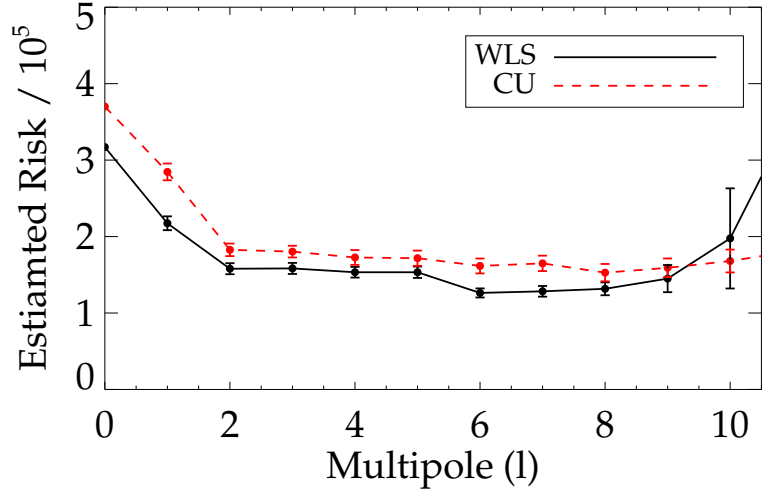}\\
\caption{Estimated risk for WLS (black-solid) and CU (red-dashed).  The estimated risk for a single $l$ is the median value from a distribution of 1000 bootstraps. As $l$ increases, the distribution becomes skewed and the estimated risk becomes unstable.  We choose the median to be robust against outliers.  This is crucial for CU as choosing many points with small sampling density in the bootstrap can make the risk very large.    The error on the estimated risk is the interquartile range (IQR) divided by 1.35 such that at low $l$ when the distribution is normal, the IQR reduces to the standard deviation.  
To determine the minimum $l$ in all estimated risk curves we choose the simplest model by finding the minimum, adding the error to the minimum, and choosing the left-most $l$ less than this value.  The minima occur at $l=6$ for WLS and $l=8$ for CU.  There is power beyond the tuning parameter for WLS and so there is a bias on the coefficients.  However, the minimum risk is lower for WLS than CU indicating that our estimate of $f(x)$ is more accurate using WLS.}
\label{fig:riskR}
\end{center}
\end{figure}

To model the peculiar velocity field with non-parametric WLS methods we first determine the tuning parameter from the estimated risk.  The risk is plotted in Figure~\ref{fig:riskR} as the solid black line.  In all estimated risk curves we determine the minimum by adding the error to the minimum risk and choosing the left-most $l$ less than this value, i.e., we choose the simplest model within the errors.  The minimum is at $l=6$; as there is power beyond this multipole (see Eq.~\ref{vTrue}), we know the coefficient estimator will be biased.

\begin{figure}[ht]
\begin{center}
\includegraphics[totalheight=4in]{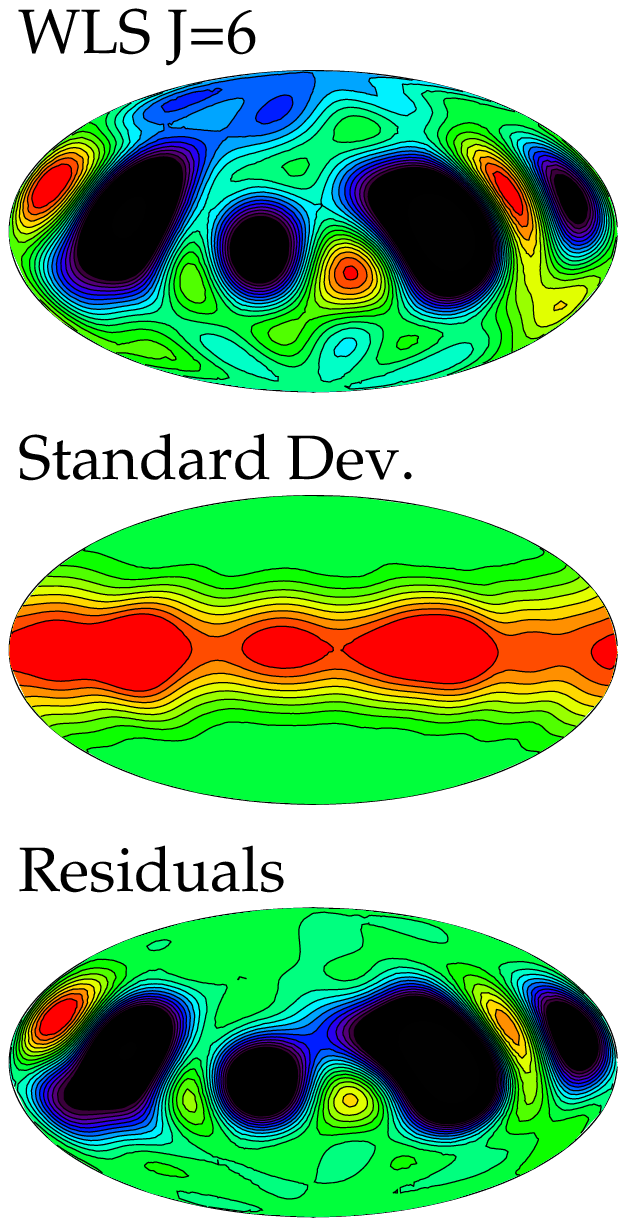}
\includegraphics[totalheight=4in]{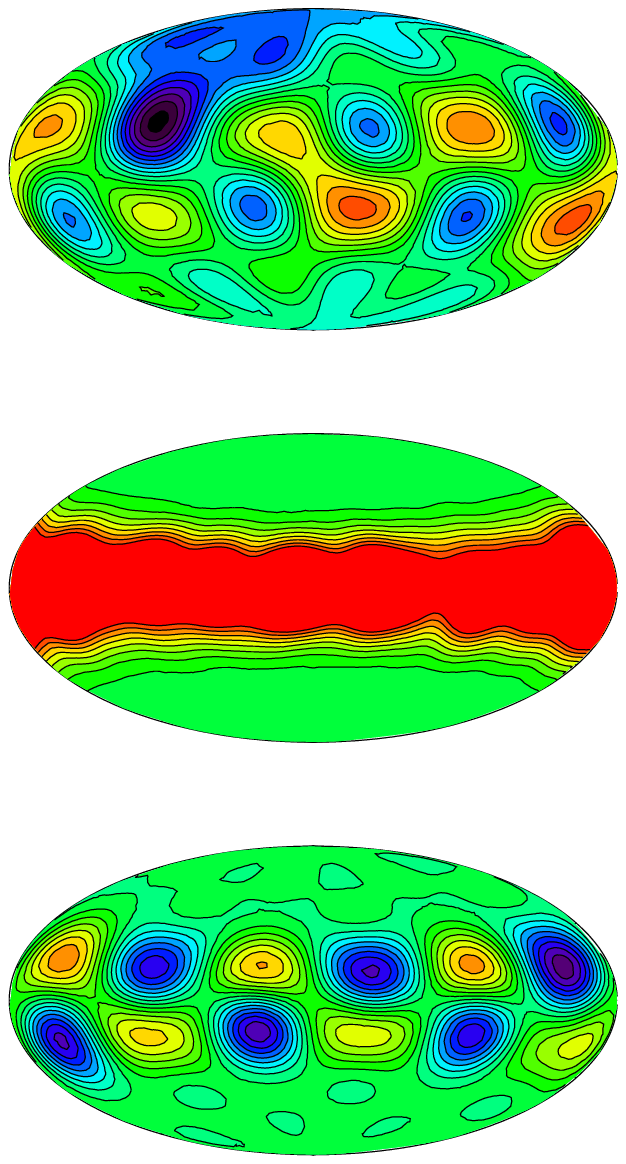}
\includegraphics[totalheight=4in]{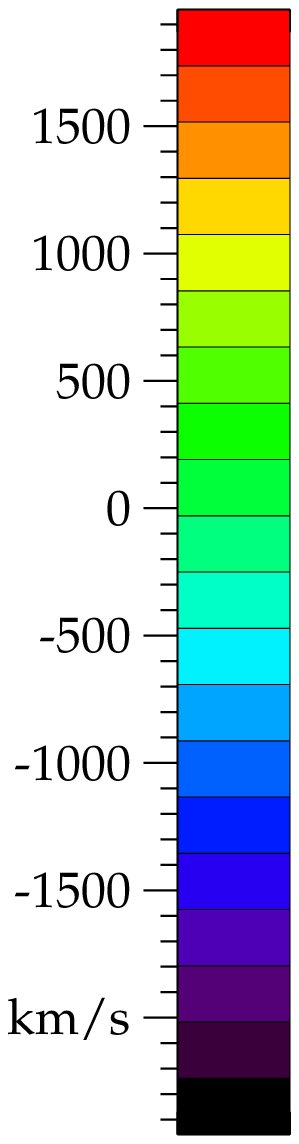}
\caption{Recovered velocity field (top), standard deviation (middle), and residuals (bottom) in km~$\rm s^{-1}$ for WLS for one realization of the data (left) and the combined results of 100 realizations of the data (right). The left plots were generated by bootstrapping a single dataset 1000 times using the tuning parameter $J=6$.  We calculate the velocity for a set of 10,000 points distributed across the sky based on the derived $a_{lm}$ coefficients for each bootstrap.  These were averaged to create a contour plot of the peculiar velocity field (top) and standard deviation (middle).  Finally, to create the residual plot, we took the difference between the averaged peculiar velocity at each point and the peculiar velocity calculated from Eq.~\ref{vTrue}. 
We perform the same analysis but combine the results of 100 realizations of the data on the right.  It is clear that the power in the galactic plane is not merely a function of one dataset but a result of power beyond the tuning parameter and the underlying sampling density.}
\label{vrecoverRH1}
\end{center}
\end{figure}

The results from WLS are in the left column in Figure~\ref{vrecoverRH1}.  The effects of the bias are clearly evident.   Artifacts appear in the galactic plane where we are not constrained by any data and are not accounted for by the standard deviation; it is not wide enough or deep enough. To determine if the power in the galactic plane is a consequence of a specific dataset, we perform 100 different realizations of the data.  If the artifacts are a function of a specific dataset, we would expect after doing many realizations that the combined results, plotted on the right side in Figure~\ref{vrecoverRH1}, would recover the true velocity field or that the standard deviation would be large enough to account for any discrepancies.  
The plots in the right column demonstrate that this is not merely a result of one realization of the data, but a result of the underlying sampling density and power beyond the tuning parameter.

\begin{figure}[ht]
\begin{center}
\includegraphics[totalheight=4in]{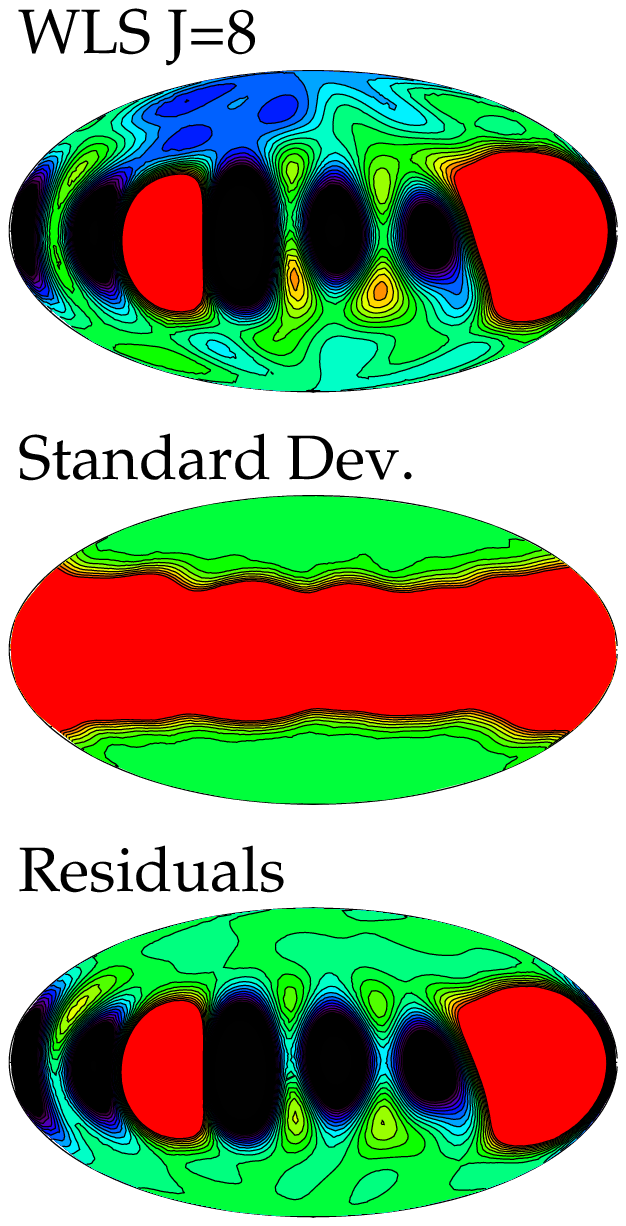}
\includegraphics[totalheight=4in]{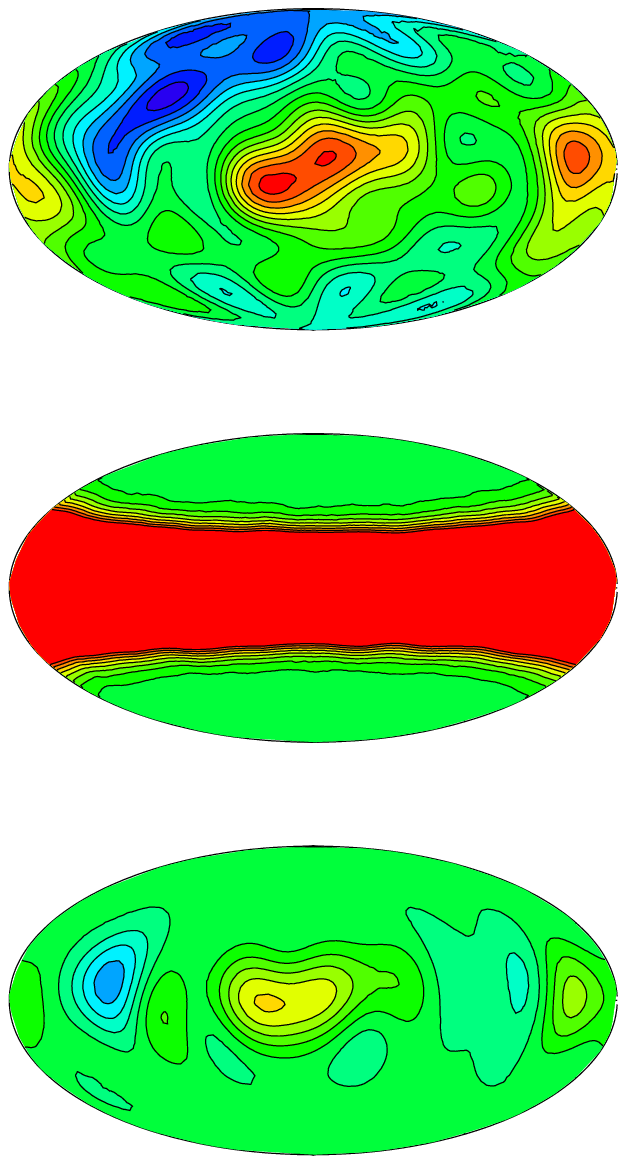}
\includegraphics[totalheight=4in]{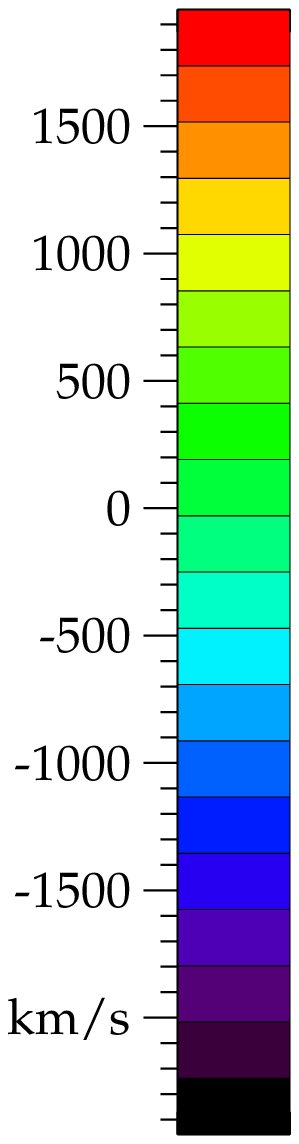}
\caption{Results for WLS forcing the tuning parameter to be $J=8$ created in an identical manner to the plots presented in Figure~\ref{vrecoverRH1}.  We see that for one realization of the data the standard deviation is sufficient to account for all of the power in the galactic plane which is not real.  The combined results from 100 realizations of the data (right) show the artifacts in the galactic plane do go away, confirming that they are due to the bias on the coefficients as a result of power beyond the tuning parameter in Figure~\ref{vrecoverRH1}.}
\label{vrecoverRH18}
\end{center}
\end{figure}

For comparison, we force the tuning parameter to be $l=8$ and perform the same analysis in Figure~\ref{vrecoverRH18}.  We confirm that there is no bias, even if the sampling density is non-uniform.  WLS now recovers the entire velocity field well and has a standard deviation large enough to account for any power fit in the galactic plane.  By combining many realizations (right) we see the anomalous power in the galactic plane average out, doing a remarkable job of recovering the true velocity field.

\subsection{Recovered Peculiar Velocity Field from CU}\label{sec:pecvel_cu}

\begin{figure}[ht]
\begin{center}
\includegraphics[totalheight=3in]{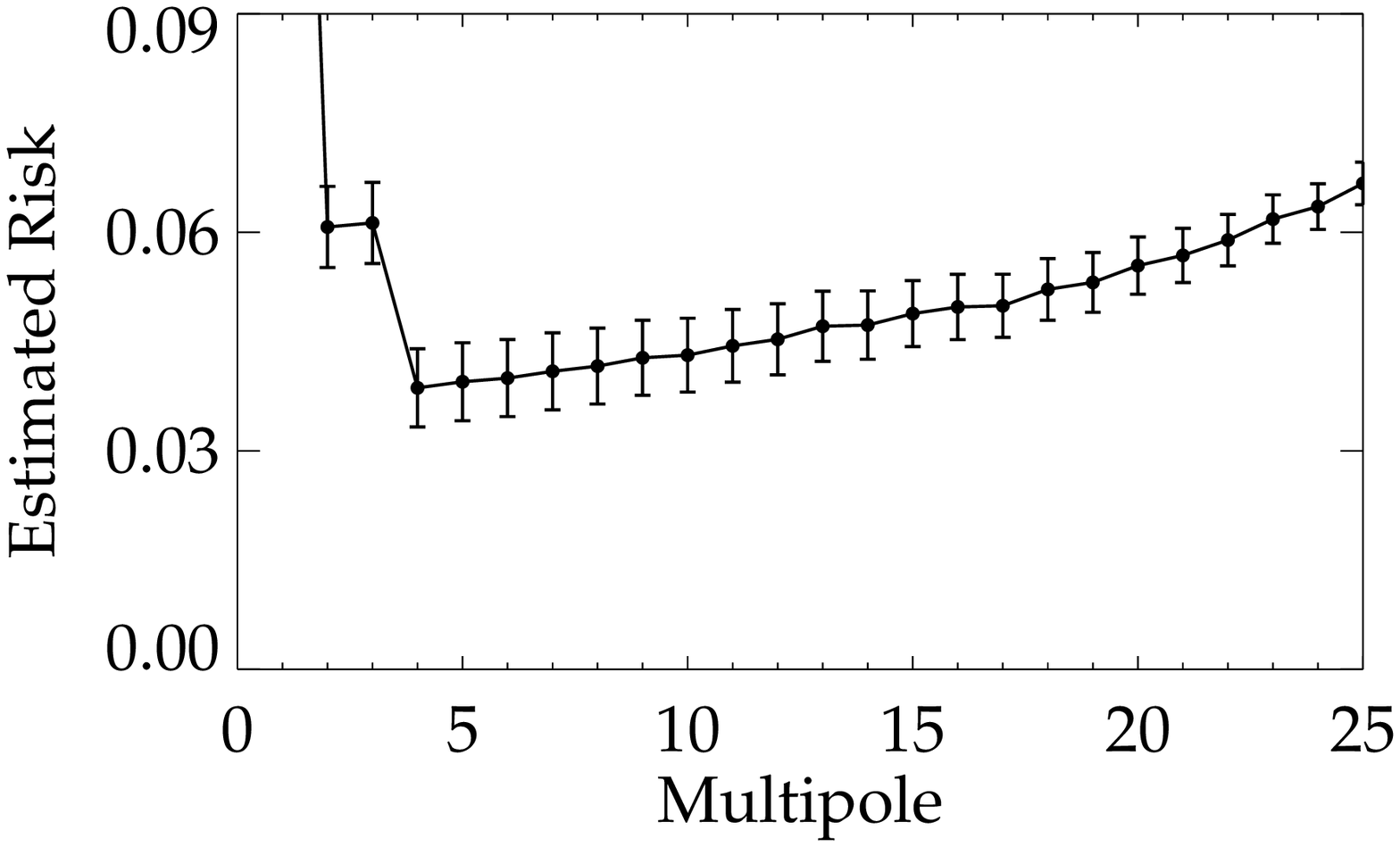}\\
\caption{Estimated risk for the sampling density with a minimum at $l=4$.  It is difficult to estimate the mean and standard deviation of the risk for the sampling density via bootstrapping because duplicates and removing points will change the inherent distribution of the data.  We therefore must use the entire dataset to estimate the risk.  This is in contrast to Fig.~\ref{fig:riskR}, where we take the median of 1000 bootstrap resamples.   The errors are estimated by dividing the data into two equal subsets, using one to calculate $Z_{i}$ and the other to calculate the estimated risk.  This is done 500 times.  The estimated errors are then the standard deviation at each $l$ scaled by 1/2 due to the decrease in the number of points used to estimate the risk.}
\label{riskH}
\end{center}
\end{figure}

\begin{figure}[ht]
\begin{center}
\includegraphics[totalheight=2in]{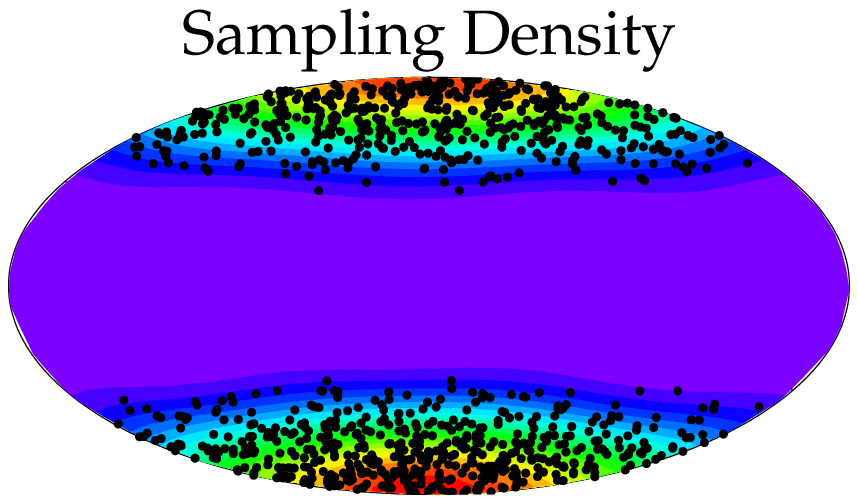}
\includegraphics[totalheight=2in]{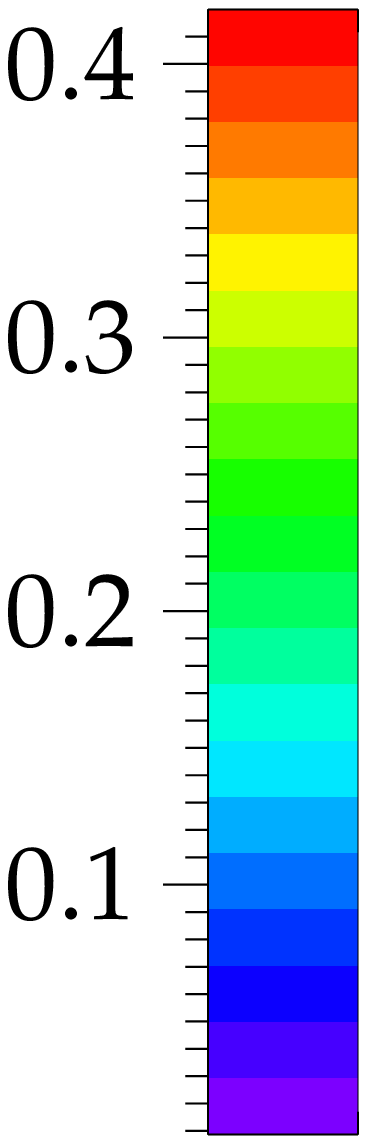}\\
\includegraphics[totalheight=2in]{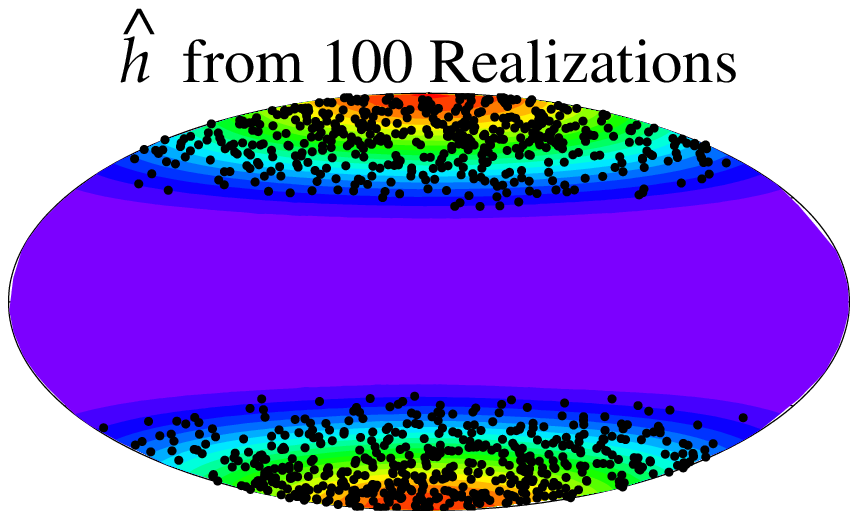}
\includegraphics[totalheight=2in]{f7b.eps}\\
\includegraphics[totalheight=2in]{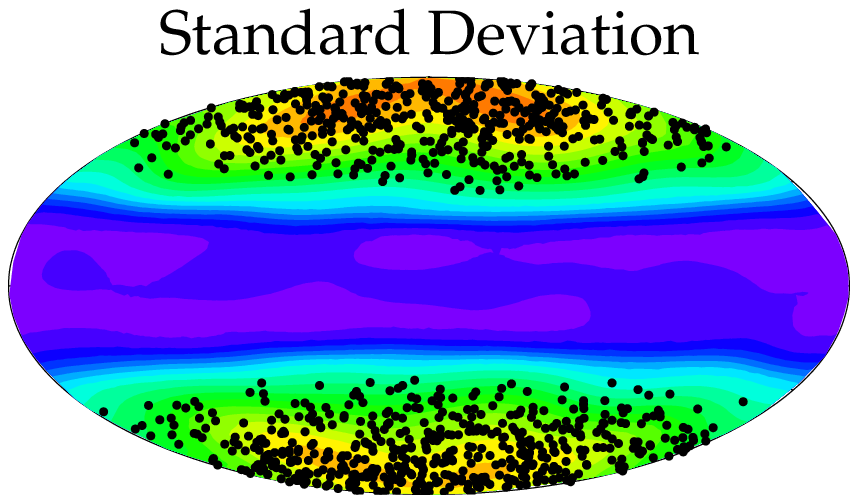}
\includegraphics[totalheight=2in]{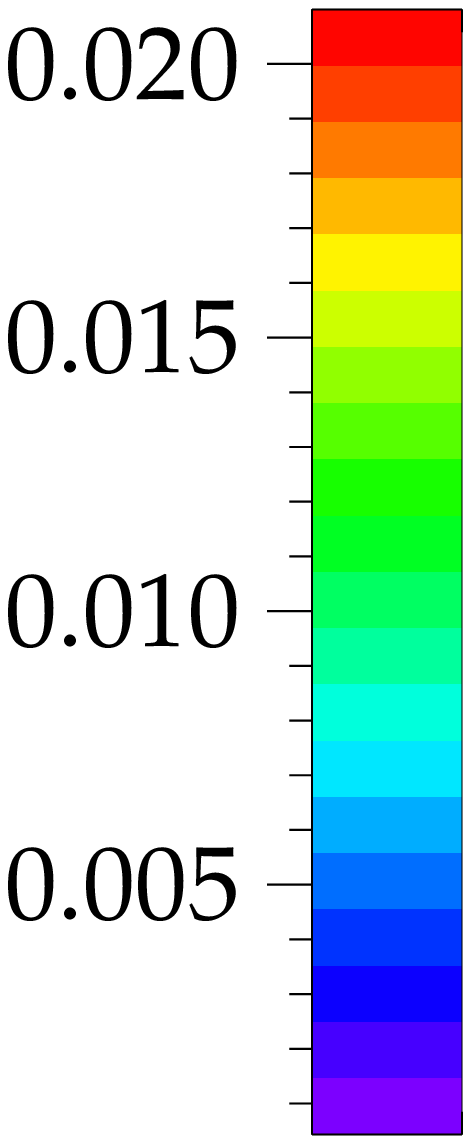}\\
\caption{Typical recovered sampling density for one realization of the data using the tuning parameter $I=4$ (top).  Over-plotted are the simulated data points.  To ensure a positive definite sampling density, we calculate $h$ according to Eq.~\ref{h}, set all negative values to zero, add a small constant to the entire field, and then renormalize. The mean (middle) and standard deviation (bottom) of the sampling density are plotted for 100 different realizations.  For each realization, the sampling density was calculated using the best tuning parameter for that dataset.  Over-plotted are the simulated data points from one realization.  The standard deviation is roughly factor of 20 lower and so $h$ is well estimated using 1000 data points. }
\label{cH}
\end{center}
\end{figure}

Removing the bias on the coefficients requires reconstructing the sampling density from the data.  The estimated risk for the sampling density is shown in Figure~\ref{riskH} with a minimum at $l=4$. Using this tuning parameter we reconstruct the sampling density according to \S\ref{sec:h}.  A contour plot of the sampling density is plotted in the top of Figure~\ref{cH} with the data points overlaid as black circles.  To investigate how well $h$ is estimated, we combine the sampling density from 100 realizations of the data and calculate the mean (middle) and standard deviation (bottom) in  Figure~\ref{cH}.  The standard deviation is about a factor of 20 smaller than the sampling density and so the sampling density is well recovered using 1000 data points. 

Having found the sampling density, we estimate the risk for CU as we did for WLS.  These results are shown in Figure~\ref{fig:riskR} (red-dashed) with a minimum at $l=8$.  Note that
the estimated risk at $l=6$ for WLS is lower than at $l=8$ for CU.  From this we expect WLS to be more accurate modeling $f(x)$ where we have data even though there is a bias on the coefficients.

\begin{figure}[ht]
\begin{center}
\includegraphics[totalheight=4in]{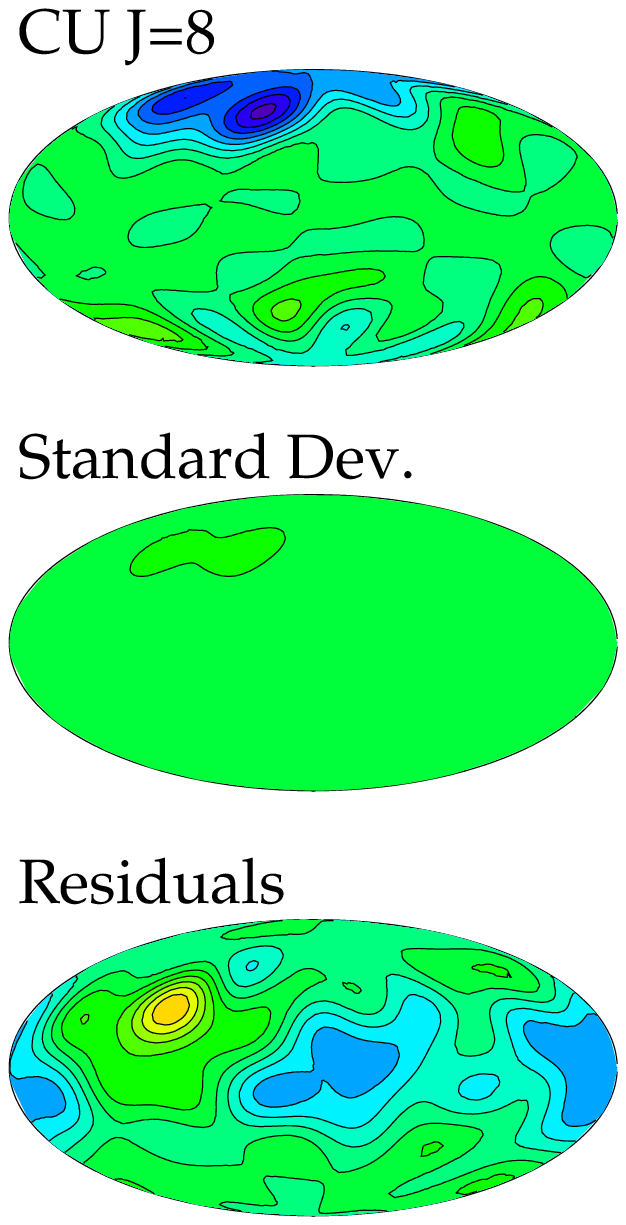}
\includegraphics[totalheight=4in]{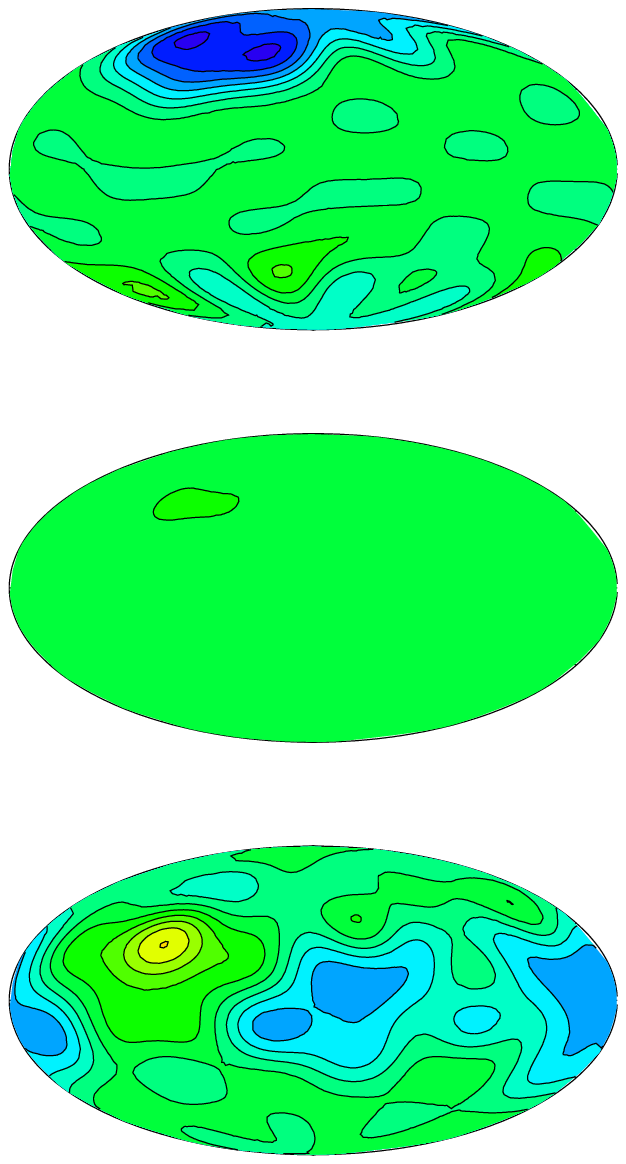}
\includegraphics[totalheight=4in]{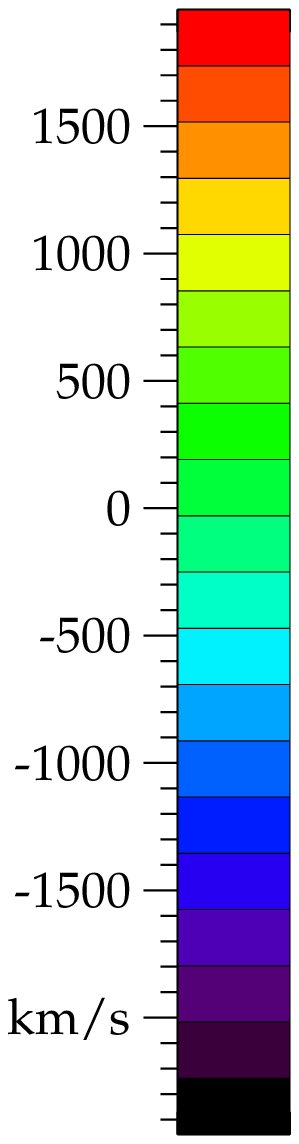}
\caption{Recovered velocity field (top), standard deviation (middle), and residuals (bottom) in km $\rm s^{-1}$ for CU for one realization of the data (left) and the combined results of 100 realizations of the data (right). These were generated in an identical manner as those in Figure~\ref{vrecoverRH1}.  By weighting by the sampling density, CU does not allow for any power in the galactic plane where there are no constraining data.}
\label{vrecoverBH1}
\end{center}
\end{figure}

\begin{figure}[ht]
\begin{center}
\includegraphics[totalheight=4in]{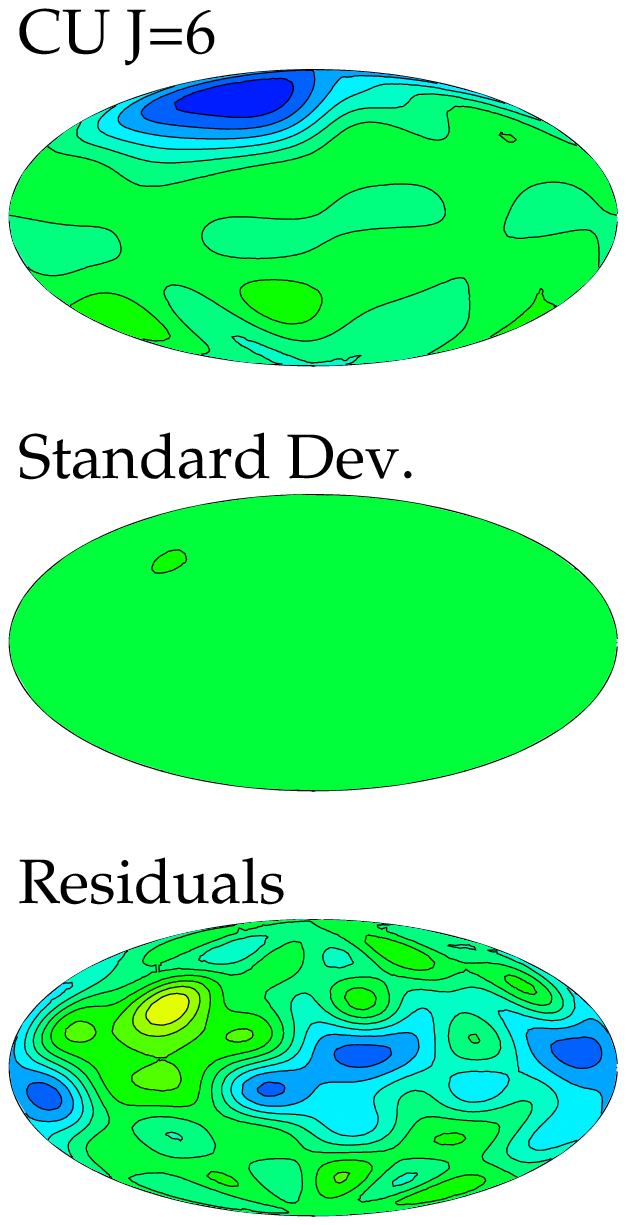}
\includegraphics[totalheight=4in]{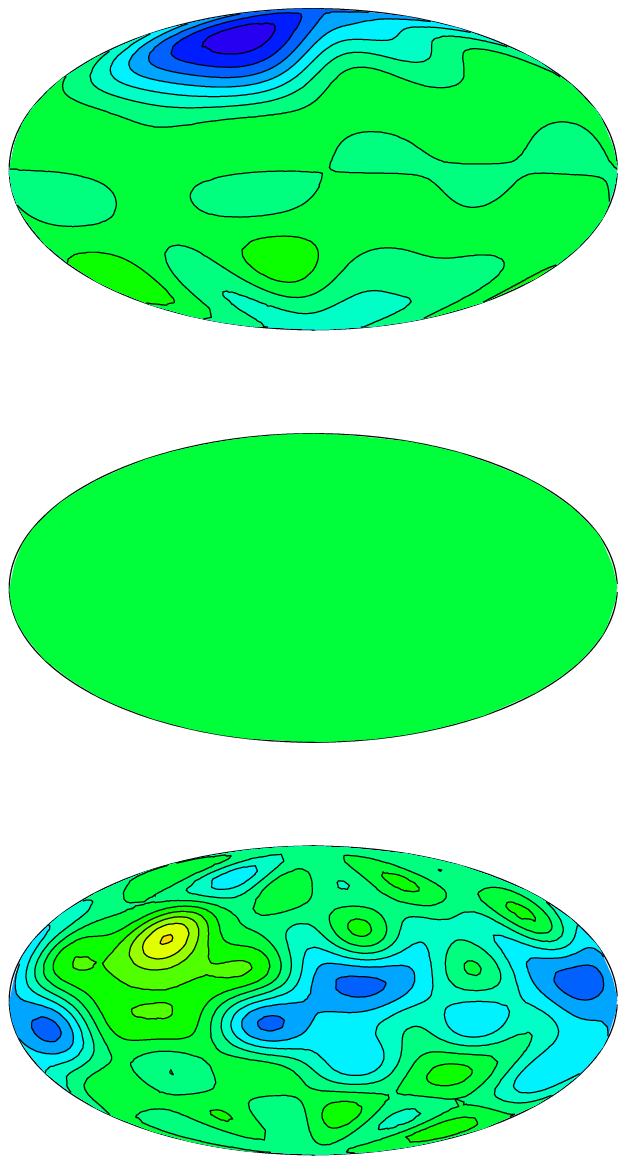}
\includegraphics[totalheight=4in]{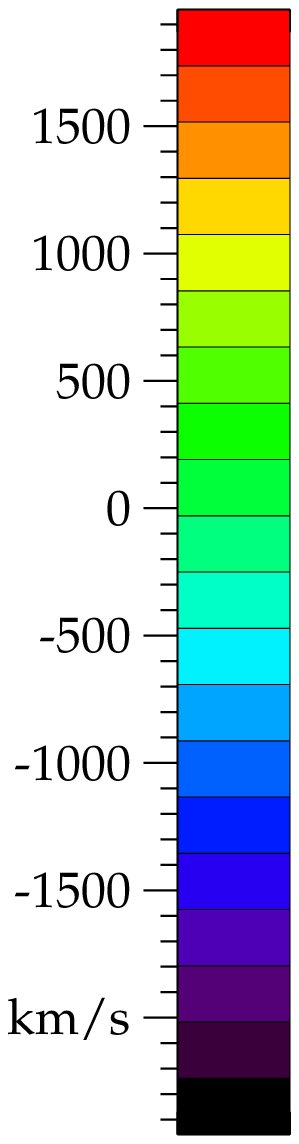}
\caption{Results for CU forcing the tuning parameter to be $J=6$. These were generated in an identical manner as those in Figure~\ref{vrecoverRH18}.  The CU method is more robust to our choice of tuning parameter.  There is power beyond $l=6$ but it is not biasing our coefficients as it did for WLS (Fig.~\ref{vrecoverRH1}).}
\label{vrecoverBH16}
\end{center}
\end{figure}

The results for CU using $J=8$ are plotted in Figure~\ref{vrecoverBH1}.  CU does not allow power to be fit in regions where there are few data points by accounting for the underlying distribution of the data.  The standard deviation also accounts for most of the discrepancies seen in the residual plot.  We also find this method to be robust against the choice of tuning parameter.  In Figure~\ref{vrecoverBH16} we force the tuning parameter to be $J=6$ and still do not see any artifacts in the galactic plane, although we have sacrificed some in overall accuracy.  This is expected since the estimated risk is larger at $J=6$.

\begin{figure}[ht]
\begin{center}
\includegraphics[totalheight=5in]{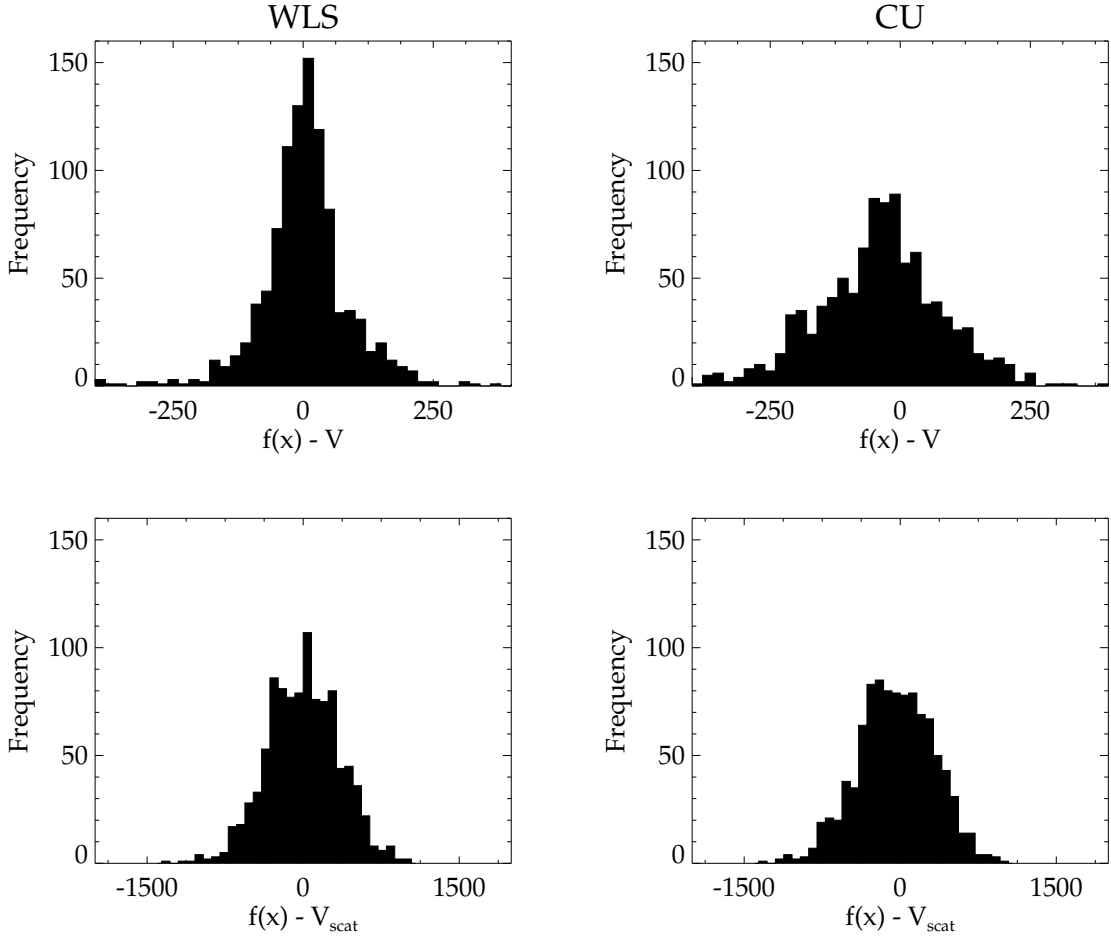}
\caption{Distributions of the residuals for each method using the tuning parameters $J_{\rm WLS}=6$ and $J_{\rm CU}=8$.  On the top row are the residuals calculated from the difference between the velocity obtained in the regression $\hat f(x)$, and the velocity $V$ given by Eq.~\ref{vTrue}.  On the bottom, the residuals are the difference between $\hat f(x)$ and the velocity scattered values $V_{\rm scat}$.  The bottom plots show us how well our methods are fitting the data.  The top plots show us how well we are recovering the true underlying peculiar velocity field where we have data.  We see that WLS models the true velocity field better as evidenced by the narrower spread in the distribution but that both methods fit the ``data'' equally well.}
\label{fighist}
\end{center}
\end{figure}

In Figure~\ref{fighist} we show distributions of the residuals for each method using the tuning parameters $J_{\rm WLS}=6$ and $J_{\rm CU}=8$.  On the top row are the residuals defined as the difference between the velocity obtained from the regression $\hat f(x)$ and the velocity $V$ given by Eq.~\ref{vTrue}.  These plots tell us how well the method is recovering the true underlying velocity field.  On the bottom, the residuals are the difference between $\hat f(x)$ and the velocity scattered values $V_{\rm scat}$.  These plots tell us how well the method is fitting simulated data.  The narrower spread in WLS in the top plots tell us it is estimating the velocity field more accurately where we have sampled.  We expect this result because the risk for WLS is lower than for CU.  Both models are fitting the simulated data similarly and have comparable spreads in their distribution (bottom plots).

\clearpage

\section{Application to Observed SNIa data}\label{sec:data}

With our framework established and tested, we now analyze real SNIa data.  We first introduce the dataset and then apply each regression method and present a
comparison of the two methods.

\subsection{SNIa Data}

Our data consist of SNIa published in \citet{Hicken09a} (hereafter H09a); \citet{JRK07, Hamuy96, Riess99} using the distance
measurements published in \citet{Hicken09b} (hereafter H09b).  Not all of the SNIa published in H09a have distance measurements published in H09b.  The rest were obtained from private communication with the author.  We use their results from the Multicolor Light Curve Shape method (MLCS2k2) \citep{JRK07} with an $R_V=1.7$ extinction law.  The positions of the SNIa are publicly available.\footnote{\url{http://www.cfa.harvard.edu/iau/lists/Supernovae.html}}  H09b provide the redshift and distance modulus $\mu$ with an assumed absolute magnitude of $M_V=-19.504$.  The peculiar velocity, $U$, is calculated according to (see \citealt{JRK07})
\begin{equation} U=H_0\,d_l(z)-H_0\,d_{\rm SN}\end{equation}
where $H_0 d_l(z)$ and $H_0 d_{\rm SN}$ are given by
\begin{equation} H_{0}\,d_{l}(z) = c (1+z) \int_{0}^{z}\left[ \Omega_M(1+{z'})^3+\Omega_{\lambda} \right]^{-1/2}d{z'}.\end{equation}
\begin{equation} H_{0}\,d_{\rm SN}=65 \left [ 10^{0.2(\mu-25)} \right]\end{equation}
Here $z$ is the redshift in the rest frame of the Local Group\footnote{\url{http://nedwww.ipac.caltech.edu/help/velc\_help.html}}
 and we assume that $\Omega_M=0.3$, $\Omega_{\lambda}=0.7$, and $H_0=65$~km~s$^{-1}$~Mpc$^{-1}$.  Our results are independent of the value we choose for $H_0$ as there is a degeneracy between $H_0$ and $M_V$. The error on the peculiar velocity is the quadrature sum of the error on $\mu$, a recommended error of 0.078~mag (see H09b), $\sigma_z$, and a peculiar velocity error of $\sigma_v=300$~km~s$^{-1}$ attributed to local motions of the SNIa which are on scales smaller than those probed in this analysis \citep{JRK07}. 

We eliminate objects which could not be fit by MLCS2k2, whose first observation occurs more than 20 days past maximum B-band light, or which showed evidence for excessive host galaxy extinction ($A_V<2$). We choose one redshift shell for our analysis due to the relatively small number of objects and consider the same velocity range adopted by \citealt{JRK07} of 1500~km~s$^{-1} \leq H_0d_{\rm SN}\leq 7500$~km~s$^{-1}$.  One object, SN~2004ap, has a particularly large peculiar velocity of 2864~km~s$^{-1}$.  Further examination reveals that this supernova, when modeled with MLCS2k2 with $R_V = 3.1$, has its first observation at 20 days past maximum B-band light.  To be conservative, we exclude this object.  This leaves us with 112 SNIa whose peculiar velocity information is recorded in Table~\ref{table:data}.  In this table we include all SNIa with $H_0d_{\rm SN} \leq 7500$~km~s$^{-1}$ for completeness.

\begin{center}
\begin{deluxetable}{lccccccccrr}
\tabletypesize{\footnotesize}
\tablewidth{0pt}
\tablecolumns{11}
\tablecaption{SNIa Data}
\label{table:data}
\tablehead{\colhead{Name}                            & 
           \colhead{RA}                              & 
	   \colhead{Dec\tablenotemark{a}}        &
	   \colhead{$z$\tablenotemark{b}}        & 
	   \colhead{$\sigma_{z}$}                    & 
	   \colhead{$\mu$}                           & 
	   \colhead{$\sigma_{\mu}$}                  & 
	   \colhead{$A_V$}                           & 
	   \colhead{$\sigma_{A_V}$}                  & 
	   \colhead{$U$}                             & 
	   \colhead{$\sigma_{U}$\tablenotemark{c}} \\
	   \colhead{}                                &  
	   \colhead{{h}\phn{m}\phn{s}}               &
	   \colhead{\phn{\arcdeg}~\phn{\arcmin}~\phn{\arcsec}}        &
	   \colhead{}               & 
	   \colhead{}               & 
	   \colhead{mag}               & 
	   \colhead{mag}               & 
	   \colhead{mag}                           & 
	   \colhead{mag}                  & 
	   \colhead{km s$^{-1}$}    & 
	   \colhead{km s$^{-1}$}    }

\startdata
1986G&13:25:36.51&-43:01:54.2&     0.003&     0.001&    28.012&     0.081&     1.221&     0.086&\ldots & \ldots\\
1990N&12:42:56.74& 13:15:24.0&     0.004&     0.001&    32.051&     0.076&     0.221&     0.051& -793&557\\
1991bg&12:25:03.71& 12:52:15.8&     0.005&     0.001&    31.728&     0.063&     0.096&     0.057&\ldots & \ldots\\
1991T&12:34:10.21& 02:39:56.6&     0.007&     0.001&    30.787&     0.062&     0.302&     0.039&\ldots & \ldots\\
1992A&03:36:27.43&-34:57:31.5&     0.006&     0.001&    31.540&     0.072&     0.014&     0.014&\ldots & \ldots\\
1992ag&13:24:10.12&-23:52:39.3&     0.026&     0.001&    35.213&     0.118&     0.312&     0.081&  224&612\\
1992al&20:45:56.49&-51:23:40.0&     0.014&     0.001&    33.964&     0.082&     0.033&     0.027&  337&458\\
1992bc&03:05:17.28&-39:33:39.7&     0.020&     0.001&    34.796&     0.061&     0.012&     0.012&  106&505\\
1992bo&01:21:58.44&-34:12:43.5&     0.018&     0.001&    34.671&     0.100&     0.034&     0.029&  122&548\\
1993H&13:52:50.34&-30:42:23.3&     0.025&     0.001&    35.078&     0.102&     0.029&     0.026&  353&556\\
1994ae&10:47:01.95& 17:16:31.0&     0.005&     0.001&    32.508&     0.067&     0.049&     0.032& -872&541\\
1994D&12:34:02.45& 07:42:04.7&     0.003&     0.001&    30.916&     0.068&     0.009&     0.009&\ldots & \ldots\\
1994M&12:31:08.61& 00:36:19.9&     0.024&     0.001&    35.228&     0.104&     0.080&     0.055& -317&606\\
1994S&12:31:21.86& 29:08:04.2&     0.016&     0.001&    34.312&     0.085&     0.047&     0.034& -190&491\\
1995ak&02:45:48.83& 03:13:50.1&     0.022&     0.001&    34.896&     0.105&     0.259&     0.072&  806&549\\
1995al&09:50:55.97& 33:33:09.4&     0.006&     0.001&    32.658&     0.074&     0.177&     0.049& -748&542\\
1995bd&04:45:21.24& 11:04:02.5&     0.014&     0.001&    34.062&     0.120&     0.462&     0.159&  183&501\\
1995D&09:40:54.75& 05:08:26.2&     0.008&     0.001&    32.748&     0.073&     0.068&     0.044& -513&439\\
1995E&07:51:56.75& 73:00:34.6&     0.012&     0.001&    33.888&     0.092&     1.460&     0.064& -211&520\\
1996ai&13:10:58.13& 37:03:35.4&     0.004&     0.001&    31.605&     0.083&     3.134&     0.056&\ldots & \ldots\\
1996bk&13:46:57.98& 60:58:12.9&     0.007&     0.001&    32.393&     0.108&     0.260&     0.098&  208&414\\
1996bo&01:48:22.80& 11:31:15.8&     0.016&     0.001&    34.305&     0.096&     0.626&     0.071&  674&492\\
1996X&13:18:01.13&-26:50:45.3&     0.008&     0.001&    32.341&     0.070&     0.031&     0.024&  -80&359\\
1997bp&12:46:53.75&-11:38:33.2&     0.009&     0.001&    32.923&     0.068&     0.479&     0.048& -196&395\\
1997bq&10:17:05.33& 73:23:02.1&     0.009&     0.001&    33.483&     0.102&     0.380&     0.055& -257&520\\
1997br&13:20:42.40&-22:02:12.3&     0.008&     0.001&    32.467&     0.067&     0.549&     0.054& -124&371\\
1997do&07:26:42.50& 47:05:36.0&     0.010&     0.001&    33.580&     0.096&     0.262&     0.061& -263&496\\
1997dt&23:00:02.93& 15:58:50.9&     0.006&     0.001&    33.257&     0.115&     1.138&     0.074& -445&702\\
1997E&06:47:38.10& 74:29:51.0&     0.013&     0.001&    34.102&     0.090&     0.085&     0.051&  -79&517\\
1997Y&12:45:31.40& 54:44:17.0&     0.017&     0.001&    34.550&     0.096&     0.096&     0.050& -298&544\\
1998ab&12:48:47.24& 41:55:28.3&     0.028&     0.001&    35.268&     0.088&     0.268&     0.047& 1009&549\\
1998aq&11:56:26.00& 55:07:38.8&     0.004&     0.001&    31.909&     0.054&     0.011&     0.011& -292&498\\
1998bp&17:54:50.71& 18:19:49.3&     0.010&     0.001&    33.175&     0.065&     0.025&     0.020&  545&412\\
1998bu&10:46:46.03& 11:50:07.1&     0.004&     0.001&    30.595&     0.061&     0.631&     0.040&\ldots & \ldots\\
1998co&21:47:36.45&-13:10:42.3&     0.017&     0.001&    34.476&     0.119&     0.123&     0.087&  548&543\\
1998de&00:48:06.88& 27:37:28.5&     0.016&     0.001&    34.464&     0.063&     0.142&     0.061&  225&519\\
1998dh&23:14:40.31& 04:32:14.1&     0.008&     0.001&    32.962&     0.090&     0.259&     0.060&  371&489\\
1998ec&06:53:06.11& 50:02:22.1&     0.020&     0.001&    34.468&     0.084&     0.041&     0.036& 1042&450\\
1998ef&01:03:26.87& 32:14:12.4&     0.017&     0.001&    34.095&     0.104&     0.068&     0.050& 1339&446\\
1998es&01:37:17.50& 05:52:50.3&     0.010&     0.001&    33.220&     0.063&     0.207&     0.042&  475&444\\
1998V&18:22:37.40& 15:42:08.4&     0.017&     0.001&    34.354&     0.090&     0.145&     0.071&  721&480\\
1999aa&08:27:42.03& 21:29:14.8&     0.015&     0.001&    34.426&     0.052&     0.025&     0.021& -701&512\\
1999ac&16:07:15.01& 07:58:20.4&     0.010&     0.001&    33.320&     0.068&     0.244&     0.042&  -78&457\\
1999by&09:21:52.07& 51:00:06.6&     0.003&     0.001&    31.017&     0.053&     0.030&     0.022&\ldots & \ldots\\
1999cl&12:31:56.01& 14:25:35.3&     0.009&     0.001&    30.945&     0.079&     2.198&     0.066&\ldots & \ldots\\
1999cp&14:06:31.30&-05:26:49.0&     0.010&     0.001&    33.441&     0.108&     0.057&     0.045& -410&475\\
1999cw&00:20:01.46&-06:20:03.6&     0.011&     0.001&    32.753&     0.105&     0.330&     0.076& 1599&322\\
1999da&17:35:22.96& 60:48:49.3&     0.013&     0.001&    33.926&     0.067&     0.066&     0.049&  136&488\\
1999dk&01:31:26.92& 14:17:05.7&     0.014&     0.001&    34.161&     0.076&     0.252&     0.058&  278&503\\
1999dq&02:33:59.68& 20:58:30.4&     0.014&     0.001&    33.705&     0.062&     0.299&     0.051&  893&411\\
1999ee&22:16:10.00&-36:50:39.7&     0.011&     0.001&    33.571&     0.058&     0.643&     0.041&  130&476\\
1999ek&05:36:31.60& 16:38:17.8&     0.018&     0.001&    34.379&     0.125&     0.312&     0.156&  406&516\\
1999gd&08:38:24.61& 25:45:33.1&     0.019&     0.001&    34.970&     0.102&     0.842&     0.066& -872&607\\
2000ca&13:35:22.98&-34:09:37.0&     0.024&     0.001&    35.182&     0.071&     0.017&     0.015&  -98&537\\
2000cn&17:57:40.42& 27:49:58.1&     0.023&     0.001&    35.057&     0.085&     0.071&     0.060&  717&543\\
2000cx&01:24:46.15& 09:30:30.9&     0.007&     0.001&    32.554&     0.067&     0.006&     0.005&  446&444\\
2000dk&01:07:23.52& 32:24:23.2&     0.016&     0.001&    34.333&     0.084&     0.017&     0.015&  745&486\\
2000E&20:37:13.77& 66:05:50.2&     0.004&     0.001&    31.788&     0.102&     0.466&     0.122&\ldots & \ldots\\
2000fa&07:15:29.88& 23:25:42.4&     0.022&     0.001&    34.987&     0.104&     0.287&     0.056&  -43&573\\
2001bf&18:01:33.99& 26:15:02.3&     0.015&     0.001&    34.059&     0.086&     0.170&     0.068&  737&452\\
2001bt&19:13:46.75&-59:17:22.8&     0.014&     0.001&    34.025&     0.089&     0.426&     0.063&  158&468\\
2001cp&17:11:02.58& 05:50:26.8&     0.022&     0.001&    34.998&     0.190&     0.054&     0.047&  448&741\\
2001cz&12:47:30.17&-39:34:48.1&     0.016&     0.001&    34.260&     0.088&     0.200&     0.070& -237&475\\
2001el&03:44:30.57&-44:38:23.7&     0.004&     0.001&    31.625&     0.073&     0.500&     0.044&\ldots & \ldots\\
2001ep&04:57:00.26&-04:45:40.2&     0.013&     0.001&    33.893&     0.085&     0.259&     0.054&  -67&478\\
2001fe&09:37:57.10& 25:29:41.3&     0.014&     0.001&    34.102&     0.092&     0.099&     0.049& -349&490\\
2001fh&21:20:42.50& 44:23:53.2&     0.011&     0.001&    33.778&     0.109&     0.077&     0.062&  335&515\\
2001G&09:09:33.18& 50:16:51.3&     0.017&     0.001&    34.482&     0.089&     0.050&     0.035&   08&506\\
2001v&11:57:24.93& 25:12:09.0&     0.016&     0.001&    34.047&     0.067&     0.171&     0.041&  349&418\\
2002bo&10:18:06.51& 21:49:41.7&     0.005&     0.001&    32.185&     0.077&     0.908&     0.050& -579&475\\
2002cd&20:23:34.42& 58:20:47.4&     0.010&     0.001&    33.605&     0.110&     1.026&     0.132&   04&544\\
2002cr&14:06:37.59&-05:26:21.9&     0.010&     0.001&    33.458&     0.085&     0.122&     0.063& -465&472\\
2002dj&13:13:00.34&-19:31:08.7&     0.010&     0.001&    33.104&     0.094&     0.342&     0.078&  -93&401\\
2002do&19:56:12.88& 40:26:10.8&     0.015&     0.001&    34.340&     0.110&     0.034&     0.034&  336&539\\
2002dp&23:28:30.12& 22:25:38.8&     0.010&     0.001&    33.565&     0.091&     0.268&     0.090&  449&490\\
2002er&17:11:29.88& 07:59:44.8&     0.009&     0.001&    32.998&     0.083&     0.227&     0.074&   99&452\\
2002fk&03:22:05.71&-15:24:03.2&     0.007&     0.001&    32.616&     0.073&     0.034&     0.023&   50&452\\
2002ha&20:47:18.58& 00:18:45.6&     0.013&     0.001&    34.013&     0.086&     0.042&     0.032&  450&490\\
2002he&08:19:58.83& 62:49:13.2&     0.025&     0.001&    35.250&     0.131&     0.031&     0.026&  317&662\\
2002hw&00:06:49.06& 08:37:48.5&     0.016&     0.001&    34.330&     0.095&     0.605&     0.099&  754&497\\
2002jy&01:21:16.27& 40:29:55.3&     0.020&     0.001&    35.188&     0.079&     0.103&     0.056& -441&620\\
2002kf&06:37:15.31& 49:51:10.2&     0.020&     0.001&    34.978&     0.089&     0.030&     0.025& -468&587\\
2003cg&10:14:15.97& 03:28:02.5&     0.005&     0.001&    31.745&     0.085&     2.209&     0.053&\ldots & \ldots\\
2003du&14:34:35.80& 59:20:03.8&     0.007&     0.001&    33.041&     0.062&     0.032&     0.022& -558&579\\
2003it&00:05:48.47& 27:27:09.6&     0.024&     0.001&    35.282&     0.120&     0.083&     0.055&  548&657\\
2003kf&06:04:35.42&-12:37:42.8&     0.008&     0.001&    32.765&     0.093&     0.114&     0.080& -267&447\\
2003W&09:46:49.48& 16:02:37.6&     0.021&     0.001&    34.867&     0.077&     0.330&     0.050& -157&516\\
2004ap&10:05:43.81& 10:16:17.1&     0.025&     0.001&    34.093&     0.174&     0.375&     0.088&\ldots & \ldots\\
2004bg&11:21:01.53& 21:20:23.4&     0.022&     0.001&    35.096&     0.096&     0.067&     0.052& -553&588\\
2004fu&20:35:11.54& 64:48:25.7&     0.009&     0.001&    33.137&     0.197&     0.175&     0.123&  336&524\\
2005am&09:16:12.47&-16:18:16.0&     0.009&     0.001&    32.556&     0.097&     0.037&     0.033&  161&337\\
2005cf&15:21:32.21&-07:24:47.5&     0.007&     0.001&    32.582&     0.079&     0.208&     0.070& -250&446\\
2005el&05:11:48.72& 05:11:39.4&     0.015&     0.001&    34.243&     0.081&     0.012&     0.013& -156&501\\
2005hk&00:27:50.87&-01:11:52.5&     0.012&     0.001&    34.505&     0.070&     0.810&     0.044&-1093&672\\
2005kc&22:34:07.34& 05:34:06.3&     0.014&     0.001&    34.084&     0.090&     0.624&     0.074&  527&498\\
2005ke&03:35:04.35&-24:56:38.8&     0.004&     0.001&    31.920&     0.054&     0.068&     0.040& -194&500\\
2005ki&10:40:28.22& 09:12:08.4&     0.021&     0.001&    34.804&     0.088&     0.018&     0.015& -138&519\\
2005ls&02:54:15.97& 42:43:29.8&     0.021&     0.001&    34.695&     0.094&     0.750&     0.064&  980&505\\
2005mz&03:19:49.88& 41:30:18.6&     0.017&     0.001&    34.298&     0.087&     0.266&     0.089&  796&468\\
2006ac&12:41:44.86& 35:04:07.1&     0.024&     0.001&    35.256&     0.091&     0.104&     0.047& -360&599\\
2006ax&11:24:03.46&-12:17:29.2&     0.018&     0.001&    34.594&     0.067&     0.038&     0.029& -542&497\\
2006cm&21:20:17.46&-01:41:02.7&     0.015&     0.001&    34.578&     0.115&     1.829&     0.079& -199&607\\
2006cp&12:19:14.89& 22:25:38.2&     0.023&     0.001&    35.006&     0.101&     0.440&     0.064&  207&554\\
2006d&12:52:33.94&-09:46:30.8&     0.010&     0.001&    33.027&     0.089&     0.076&     0.042& -214&409\\
2006et&00:42:45.82&-23:33:30.4&     0.021&     0.001&    35.065&     0.112&     0.328&     0.074&  172&614\\
2006eu&20:02:51.15& 49:19:02.3&     0.023&     0.001&    34.465&     0.141&     1.208&     0.119& 2423&492\\
2006h&03:26:01.49& 40:41:42.5&     0.014&     0.001&    34.259&     0.084&     0.287&     0.125& -207&545\\
2006ke&05:52:37.38& 66:49:00.5&     0.017&     0.001&    34.984&     0.128&     1.006&     0.203&-1068&698\\
2006kf&03:41:50.48& 08:09:25.0&     0.021&     0.001&    34.961&     0.113&     0.024&     0.024&  135&596\\
2006le&05:00:41.99& 63:15:19.0&     0.017&     0.001&    34.633&     0.092&     0.076&     0.060&  -03&545\\
2006lf&04:38:29.49& 44:02:01.5&     0.013&     0.001&    33.745&     0.123&     0.095&     0.074&  487&468\\
2006mp&17:12:00.20& 46:33:20.8&     0.023&     0.001&    35.259&     0.104&     0.166&     0.068&  -69&633\\
2006n&06:08:31.24& 64:43:25.1&     0.014&     0.001&    34.174&     0.083&     0.027&     0.023&   53&500\\
2006sr&00:03:35.02& 23:11:46.2&     0.023&     0.001&    35.280&     0.098&     0.085&     0.053&  305&624\\
2006td&01:58:15.76& 36:20:57.7&     0.015&     0.001&    34.464&     0.136&     0.171&     0.079&  -56&606\\
2006x&12:22:53.99& 15:48:33.1&     0.006&     0.001&    30.958&     0.077&     2.496&     0.043&\ldots & \ldots\\
2007af&14:22:21.06& 00:23:37.7&     0.006&     0.001&    32.302&     0.082&     0.215&     0.054& -303&433\\
2007au&07:11:46.11& 49:51:13.4&     0.020&     0.001&    34.624&     0.081&     0.049&     0.039&  667&479\\
2007bc&11:19:14.57& 20:48:32.5&     0.022&     0.001&    34.932&     0.108&     0.084&     0.059&  -64&564\\
2007bm&11:25:02.30&-09:47:53.8&     0.007&     0.001&    32.382&     0.101&     0.975&     0.073& -320&390\\
2007ca&13:31:05.81&-15:06:06.6&     0.015&     0.001&    34.622&     0.096&     0.580&     0.069&-1337&599\\
2007ci&11:45:45.85& 19:46:13.9&     0.019&     0.001&    34.290&     0.090&     0.074&     0.063&  690&434\\
2007cq&22:14:40.43& 05:04:48.9&     0.025&     0.001&    35.085&     0.101&     0.109&     0.059& 1399&558\\
2007s&10:00:31.26& 04:24:26.2&     0.014&     0.001&    34.222&     0.074&     0.833&     0.054& -942&523\\
2008bf&12:04:02.90& 20:14:42.6&     0.025&     0.001&    35.174&     0.078&     0.102&     0.049&  271&535\\
2008L&03:17:16.65& 41:22:57.6&     0.019&     0.001&    34.392&     0.193&     0.036&     0.033& 1117&602\\
\enddata
\tablenotetext{a}{SNIa RA and Dec [J2000] from \url{http://www.cfa.harvard.edu/iau/lists/Supernovae.html}}
\tablenotetext{b}{Redshift in the rest frame of the Cosmic Microwave Background}
\tablenotetext{c}{Includes the error on $\mu$, a recommended error of 0.078 mag (see H09), $\sigma_z$, and a peculiar velocity error of $\sigma_v=300$~km s$^{-1}$ due to local motions on scales smaller than those probed by this analysis.}
\end{deluxetable}
\end{center}

\begin{figure}[ht]
\begin{center}
\includegraphics[totalheight=3in]{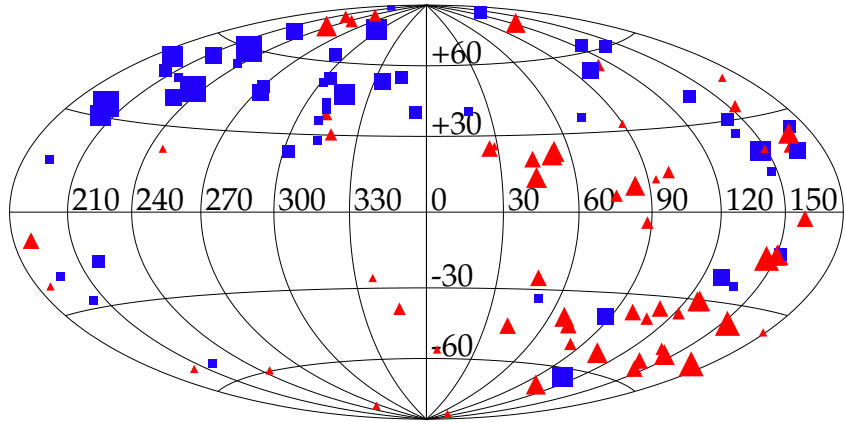}\\
\caption{Sky distribution of 112 SNIa taken from \citet{Hicken09a} in Galactic longitude and latitude.  The velocity range considered is 1500~km~s$^{-1} \leq H_0d_{\rm SN}\leq 7500$~km~s$^{-1}$ where $d_{\rm SN}$ is the luminosity distance.   The color/shape of the points indicates positive (red-triangle) and negative (blue-square) peculiar velocities and the size corresponds to the magnitude of the peculiar velocity.  From this figure one can clearly see a dipole signature between the upper-left and lower-right quadrants.}
\label{fig:data}
\end{center}
\label{fig:snriskwls}
\end{figure}

In Figure~\ref{fig:data} we plot the distribution of the data. One can clearly see the dipole with significant concentrations of negative peculiar velocities around $(l,b) = (260^{\circ}, 40^{\circ})$ and positive peculiar velocities around $(l,b)=(100^{\circ},-40^{\circ})$.  As we will explore in future figures (e.g. fig. \ref{fig:contourWLSCU}) this dipole structure is consistent with the dipole in the temperature anisotropy of the Cosmic Microwave Background.

\subsection{WLS and CU Regressions on SNIa Data}

\begin{figure}[ht]
\begin{center}
\includegraphics[totalheight=3in]{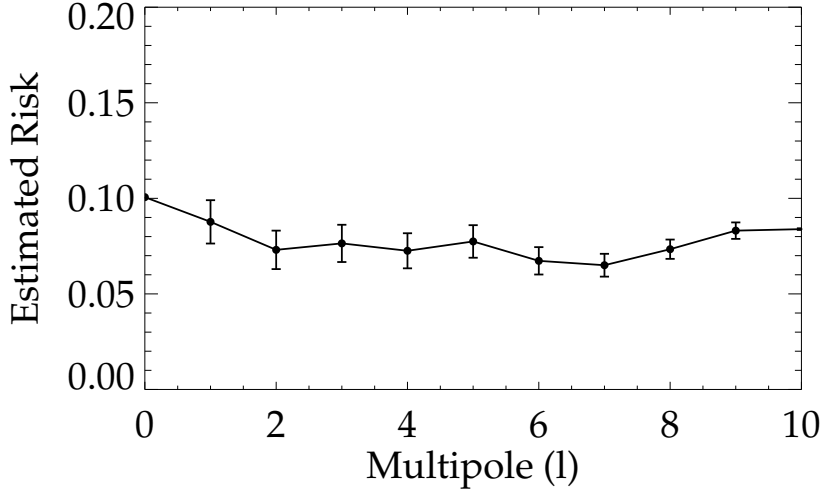}\\ 
\caption{Estimated risk for sampling density with a minimum at $l=6$ generated in a similar fashion to Fig.~\ref{riskH}.   The high $l$ moment is an indication of the lumpiness in our sampling density.}
\label{fig:riskhdata}
\end{center}
\end{figure}

\begin{figure}[ht]
\begin{center}
\includegraphics[totalheight=3in]{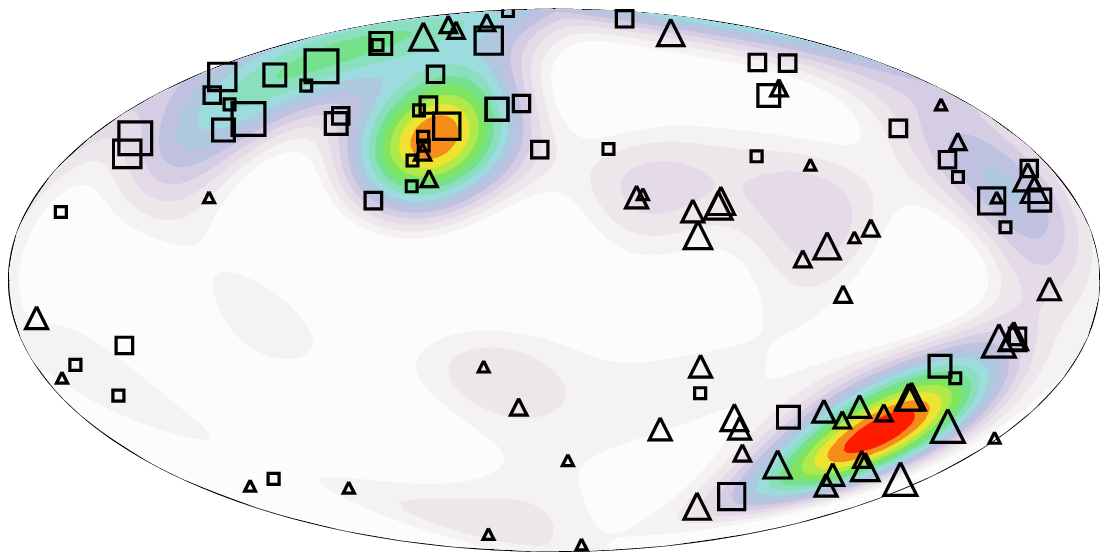} 
\includegraphics[totalheight=3in]{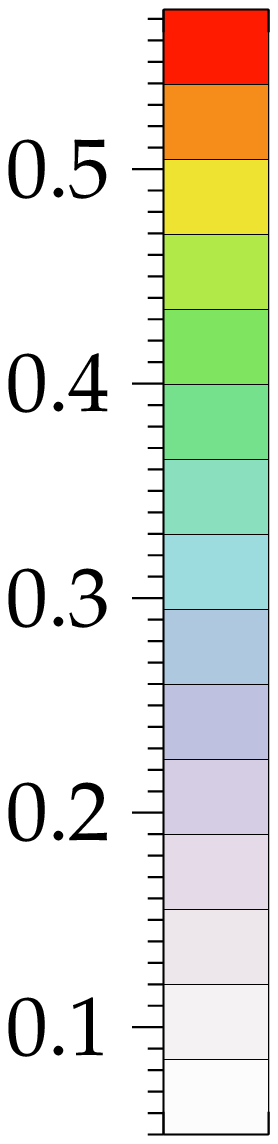}\\ 
\caption{Recovered sampling density using $I=6$ as the tuning parameter.  After calculating $h$ according to Eq.~\ref{h}, the negative values were set to zero, a small constant of 0.05 was added, and the sampling density was renormalized.  Data points residing in low sampling density regions may be an indication that spherical harmonics are not the best way to decompose $h$.  Same as Fig.~\ref{fig:snriskwls}.}
\label{fig:contourhdata}
\end{center}
\end{figure}

The first step in estimating the field with CU is to calculate the sampling density.  
We follow the procedure described in \S\ref{sec:pecvel_cu} and plot the results in Figure~\ref{fig:riskhdata}.  Choosing the simplest model gives us a tuning parameter of $I=6$.  The high $l$ moment is necessary to describe the patchiness of the data distribution.  
The sampling density field is shown in Figure~\ref{fig:contourhdata}.  While it should be unlikely that we sample many data points in regions with low sampling density, there are some regions of the sky where the sampling density is very low and we have a data point.  This discrepancy, in combination with a relatively flat estimated risk function is an indication that there are likely better basis functions than spherical harmonics to use to estimate $h$.  However, as discussed in \S\ref{sec:h} they will serve for the purposes of demonstrating our method. 

\begin{figure}[ht]
\begin{center}
\includegraphics[totalheight=3in]{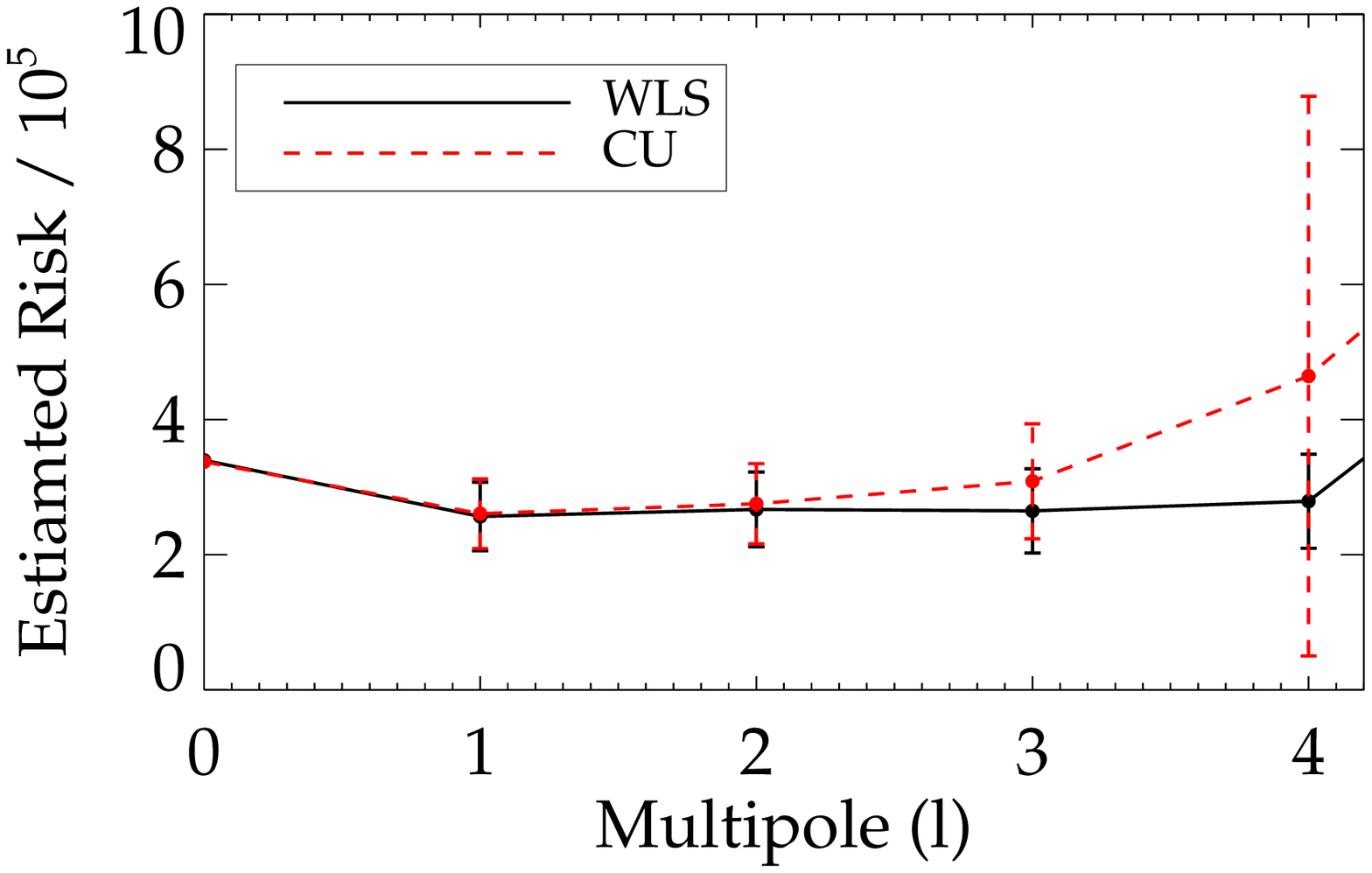}\\
\caption{Estimated risk from the mean and standard deviation of 10,000 bootstraps as a function of $l$ moment for WLS (solid-black) and CU (dashed-red) using 112 SNIa.  We find the minimum to be at $l=1$ for both methods, suggesting that the data are inadequate for detecting the quadrupole.  We expect the velocity fields derived from CU and WLS to be consistent as indicated by the similar risk values.}
\label{fig:RiskR}
\end{center}
\end{figure}

The estimated risk for CU and WLS are plotted in Figure~\ref{fig:RiskR}.  We find the tuning parameter to be $J=1$ for both methods and the risk values to be very similar.  From this we expect that the two methods will be consistent and recover the velocity field with similar accuracy. 
The current SNIa data are insufficient to detect power beyond the dipole.  Using this tuning parameter we calculate the $a_{lm}$ coefficients and the monopole and dipole terms from the following equations
\begin{eqnarray}
\rm {Monopole} &=& \frac{{a_{00}}}{\sqrt {4 \pi}}\\
\rm {Dipole}   &=& \sqrt{\frac{3}{4\pi}}\sqrt{a_{10}^2 +  \Re({a_{11}})^2 + \Im({a_{11}})^2}\\
\phi&=&-{\rm arctan} \left( \frac{\Im ({a_{11}})}{\Re({a_{11}})} \right) \\
\theta&=& {\rm arccos}\left( \frac{{a_{10}}}{\sqrt{ a_{10}^2 +  \Re({ a_{11}})^2 + \Im({ a_{11}})^2}} \right)
\end{eqnarray}
These results are summarized in Table~\ref{table:resultsWLSCU}.

\begin{center}
\begin{deluxetable}{lrrl}
\tabletypesize{\footnotesize}
\tablewidth{0pt}
\tablecolumns{4}
\tablecaption{Summary of Results}
\tablehead{\multicolumn{2}{c}{WLS} & \multicolumn{2}{c}{CU} }
\startdata
Monopole   &  149  $\pm$ 52 $\rm km$ $\rm s^{-1}$ & Monopole   &  98  $\pm$ 45 $\rm km$ $\rm s^{-1}$\\ 
Dipole     &  538 $\pm$ 86 $\rm km$ $\rm s^{-1}$ & Dipole     &  446 $\pm$ 101 $\rm km$ $\rm s^{-1}$\\ 
Galactic $l$ &  258$^{\circ}$  $\pm$ 10$^{\circ}$  & Galactic $l$ &  273$^{\circ}$  $\pm$ 11$^{\circ}$  \\ 
Galactic $b$ &  36$^{\circ}$   $\pm$ 11$^{\circ}$  &  Galactic $b$ &  46$^{\circ}$   $\pm$ 8$^{\circ}$  
\enddata
\label{table:resultsWLSCU} 
\end{deluxetable}
\end{center}

\begin{figure}[ht]
\begin{center}
\includegraphics[totalheight=2.5in]{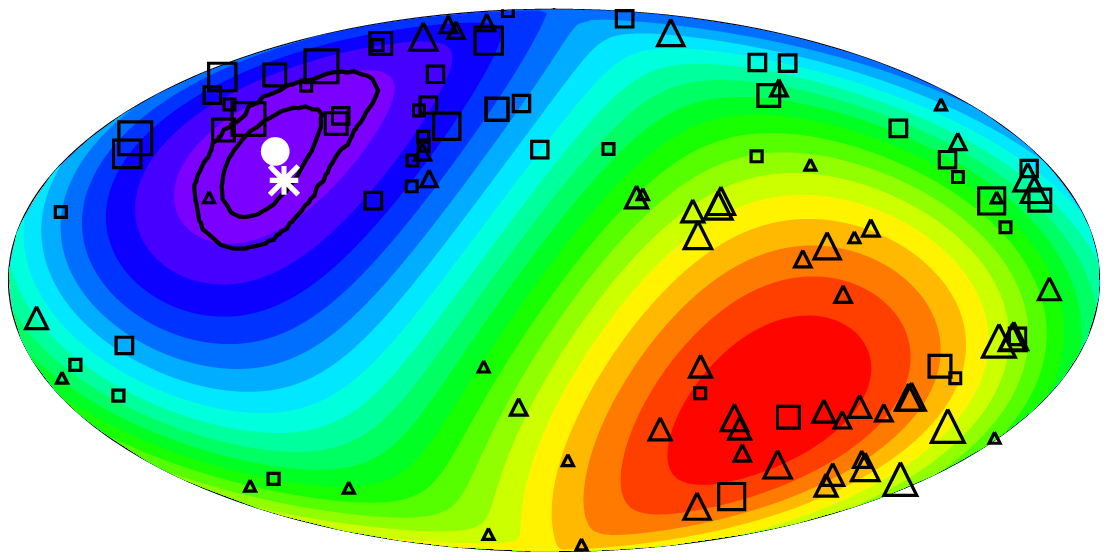}
\includegraphics[totalheight=2.5in]{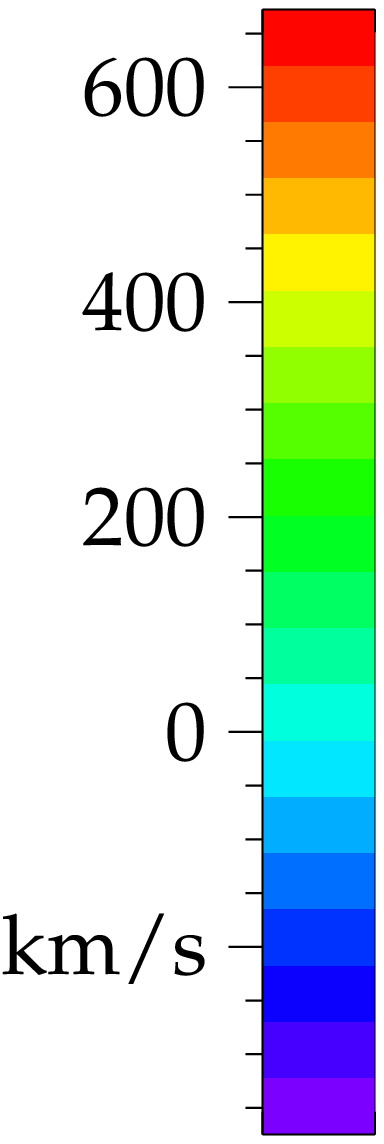}
\includegraphics[totalheight=2.5in]{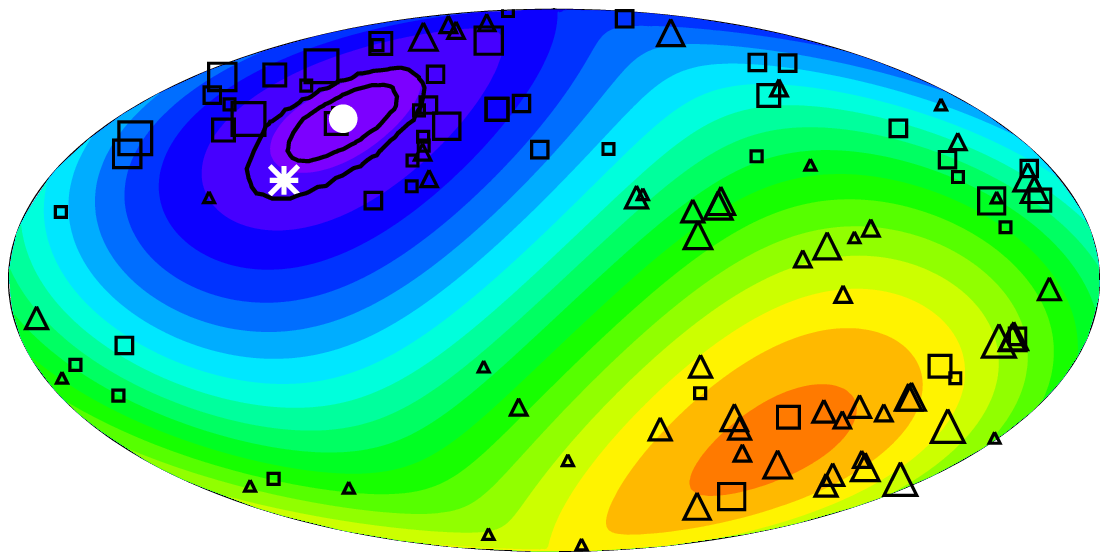}
\includegraphics[totalheight=2.5in]{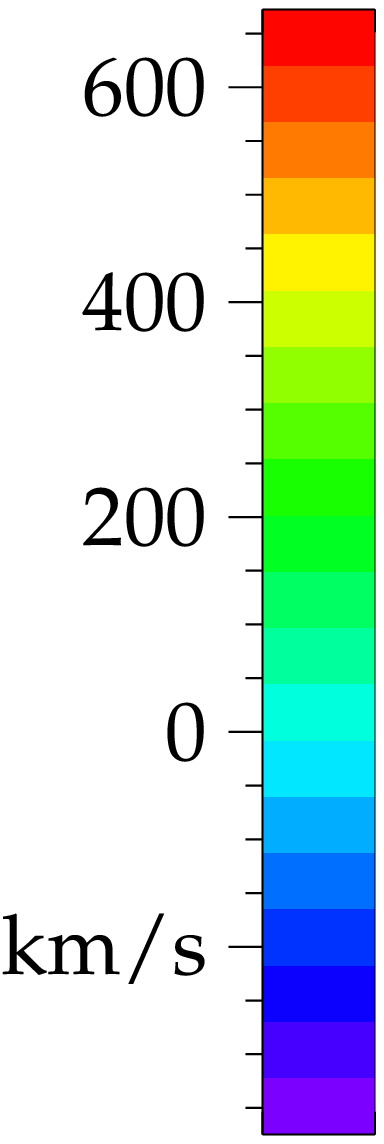}
\caption{Peculiar velocity field for WLS (top) and CU (bottom) using the tuning parameter $J=1$.  The solid white circle marks the direction of the regression dipole, the white asterisk marks the CMB dipole in the rest frame of the Local Group, and the black triangles and squares mark the data points with positive and negative peculiar velocities.  Contours are given to mark the 65\% and 95\% confidence bands of the direction of the dipole. The color scale indicates the peculiar velocity in km $\rm s^{-1}$.  We see that WLS and CU are in 95\% agreement with the direction of the CMB dipole.}
\label{fig:contourWLSCU}
\end{center}
\end{figure}

The velocity fields from WLS and CU are plotted in Figure~\ref{fig:contourWLSCU}.  The magnitudes are comparable between the two methods, with WLS being slightly larger.  The direction of the CU dipole points more toward a region of space which is well sampled.  The WLS dipole is pulled toward a less sampled region which may be why the bulk flow measurement is larger in magnitude.  This is explored more in \S\ref{sec:dipoleAnalysis}.

To compare the CU and WLS bulk motions, we use a paired t-test.   Because the coefficients were determined from the same set of data there is covariance between the parameters estimated from the two methods.  Consider bootstrapping the data N times.  For a single bootstrap let $X$ be a coefficient from CU and $Y$ be the same coefficient but derived from WLS.  The paired t-statistic comes from the distribution of $X-Y$ 
\begin{equation}
t=\frac{\ \, \langle X-Y \rangle \ \,}{\sigma_{X-Y}}
\end{equation}
where $\sigma_{X-Y}$ is the standard deviation of the $X-Y$ distribution and $\langle X-Y \rangle$ is the mean. According to the central limit theorem, for large samples many test statistics are approximately normally distributed.  For normally distributed data, $t~<$~1.96 indicates that the values being compared are in 95\% agreement. 
Performing a paired t-test on the measurements finds that the coefficients are in 95\% agreement with the results summarized in Table~\ref{table:t}.  Because the risk values are so similar (Figure~\ref{fig:RiskR}), we expect the methods to model the peculiar velocity field equally well and therefore expect the values to be statistically consistent.

\begin{center}
\begin{deluxetable}{cccc}
\tablewidth{0pt}
\tablecolumns{4}
\tablecaption{Paired t-test results}
\tablehead{\multicolumn{1}{c}{$a_{00}$} & \multicolumn{1}{c}{$a_{10}$} & \multicolumn{1}{c}{$\Re(a_{11})$} & \multicolumn{1}{c}{$\Im(a_{11})$} }
\startdata
1.50 &    0.23 &     1.72  &     1.66 \\[1ex]
\enddata
\label{table:t} 
\end{deluxetable}
\end{center}

To compare two independent measurements we perform a two-sample t-test, which gives us a statistical measure of how significant the difference between two numbers are.  We first calculate the standardized test statistic $t = (x_1 - x_2)/\sqrt{\sigma_1^2 + \sigma_2^2}$, where $x_1$ and $x_2$ are the mean values of two measurements to be compared and $\sigma_1$ and $\sigma_2$ are the associated uncertainties.  This statistic is suitable for comparing the CU or WLS bulk motion with the CMB dipole. We find the WLS Local Group bulk flow moving at $538 \pm 86$~km~s$^{-1}$ towards ($l,b$) $=$ ($258^{\circ} \pm 10^{\circ}, 36^{\circ} \pm 11^{\circ}$) which is consistent with the magnitude of the CMB dipole (635~km~s$^{-1}$) and direction ($269^{\circ}, 28^{\circ}$) with an agreement of $t_{\rm dip}=1.12$, $t_l=1.1$, and $t_{b}=0.73$.  The CU bulk flow is moving at $446 \pm 101$~km~s$^{-1}$ towards ($l,b$) $=$ ($273^{\circ} \pm 11^{\circ}, 46^{\circ} \pm 8^{\circ}$).  The CU bulk flow is in good agreement with the CMB dipole with $t_{\rm dip}=1.88$, $t_l=0.36$, and $t_b =2.25$. 

There is no strong evidence for a monopole component of the velocity field for either method. This merely demonstrates that we are using consistent values of $M_V$ and $H_0$.  For this analysis to be sensitive to a ``Hubble bubble''~\citep[e.g.,][]{JRK07}, we would look for a monopole signature as a function of redshift.  

We can directly compare our results to those obtained in \citet{JRK07} using a two-sample t-test as our analysis covers the same depth and is in the same reference frame.  They find a velocity of 541~$\pm$~75~km~s$^{-1}$ toward a direction of ($l,b$) = ($258^{\circ} \pm 18^{\circ},51^{\circ} \pm 12^{\circ}$).  Our results for WLS and CU are compatible with \citet{JRK07} with $t < 1$ in magnitude and direction.
We can also compare our results to those in \citet{Haugbolle07} for their 4500 sample transformed to the Local Group rest frame.  They find a velocity of 
516~km~s$^{-1}$ toward ($l,b$) = (248$^{\circ}$, 51$^{\circ}$). 
Their derived amplitude is slightly lower as their fit for the peculiar velocity field includes the quadrupole term.  We note from the estimated risk curves, that it is not unreasonable to fit the quadrupole as the estimated risk is similar at $l=1$ and $l=2$.  However, it is unclear if fitting the extra term improves the accuracy with which the field is modeled. Our results for CU and WLS agree with \citet{Haugbolle07} with $t \leq 1$.  Note that these $t$ values may be slightly underestimated as a subset of SNIa are common between the two analyzes.  A summary of dipole measurements is presented in Table~\ref{table:dipolesummary}.

\begin{center}
\begin{deluxetable}{lrrcll}
\tablewidth{0pt}
\tablecolumns{6}
\tablecaption{Summary of Dipole Results}
\tablehead{\multicolumn{1}{c}{Method} & \multicolumn{1}{c}{\#} & \multicolumn{1}{c}{Redshift Range} & \multicolumn{1}{c}{Depth} & \multicolumn{1}{c}{Magnitude} & \multicolumn{1}{c}{Direction}\\ 
           \multicolumn{1}{c}{} & \multicolumn{1}{c}{SNIa} & \multicolumn{1}{c}{CMB} & \multicolumn{1}{c}{km~s$^{-1}$} & \multicolumn{1}{c}{km~s$^{-1}$} & \multicolumn{1}{c}{Galactic $(l,b)$}}
\startdata
WLS                   & 112  & 0.0043-0.028   & 4000 &  $538 \pm  86$       &  $(258^\circ, 36^\circ) \pm (10^\circ, 11^\circ)$ \\ 
CU                    & 112  & 0.0043-0.028   & 4000 &  $446 \pm 101$       &  $(273^\circ, 46^\circ) \pm (11^\circ,  8^\circ)$ \\ 
\citet{Haugbolle07}   & 74   & 0.0070-0.035   & 4500 &  $516 \pm {}^{57}_{79}$  &  $(248^\circ, 51^\circ) \pm (^{15^\circ}_{20^\circ}, ^{15^\circ}_{14^\circ})$     \\
\citet{JRK07}         & 69   & 0.0043-0.028   & 3800 &  $541 \pm 75$       &  $(258^{\circ}, 51^\circ) \pm (18^{\circ}, 12^{\circ})$  \\
\enddata
\label{table:dipolesummary} 
\end{deluxetable}
\end{center}

\section{Dependence of CU and WLS on Bulk Flow Direction}\label{sec:dipoleAnalysis}

The WLS and CU analyses on real SNIa data give dipole directions that follow the well-sampled region.  This may raise suspicion that the CU method is following the sampling when determining the dipole. In this section we examine the behavior of our methods on simulated data as we vary the direction of the dipole. 

We create simulated data sets from the sampling density derived from the actual data to verify the robustness of our analysis. 
We test two randomly chosen bulk flows which vary in magnitude and direction and sample 200 SNIa for each case.  One dipole points toward a well-populated region of space and the other into a sparsely sampled region.  
A weak quadrupole is added such that the estimated risk gives a minimum at $l=1$.  There is power beyond the tuning parameter so we expect a bias to be introduced onto the coefficients for WLS.  The velocity fields for the two cases are given by
\begin{eqnarray*} 
{\rm Case~1}~:~V&=& 400Y_{01} + 590\Re{Y_{11}}+830\Im{Y_{11}} \\
&& -100Y_{20}+200\Re{Y_{21}}+250\Im{Y_{21}} -175\Re{Y_{22}} +140\Im{Y_{22}}\\
{\rm Case~2}~:~V&=& -642Y_{01} + -38\Re{Y_{11}}+810\Im{Y_{11}} \\
&& -100Y_{20}+200\Re{Y_{21}}+250\Im{Y_{21}} -175\Re{Y_{22}} +140\Im{Y_{22}}
\end{eqnarray*}

\begin{figure}[ht]
\begin{center}
\includegraphics[totalheight=3.5in]{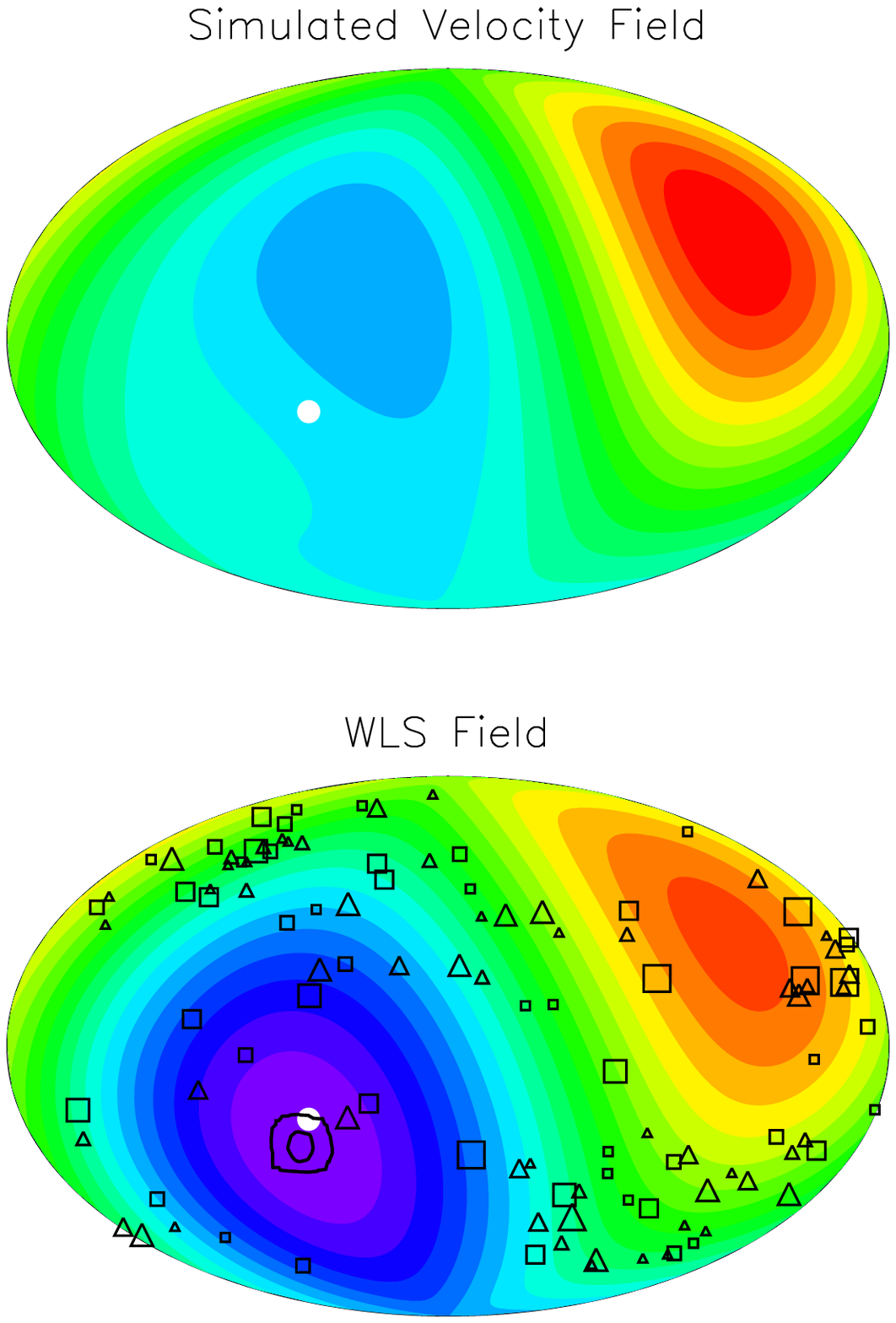}
\includegraphics[totalheight=3.5in]{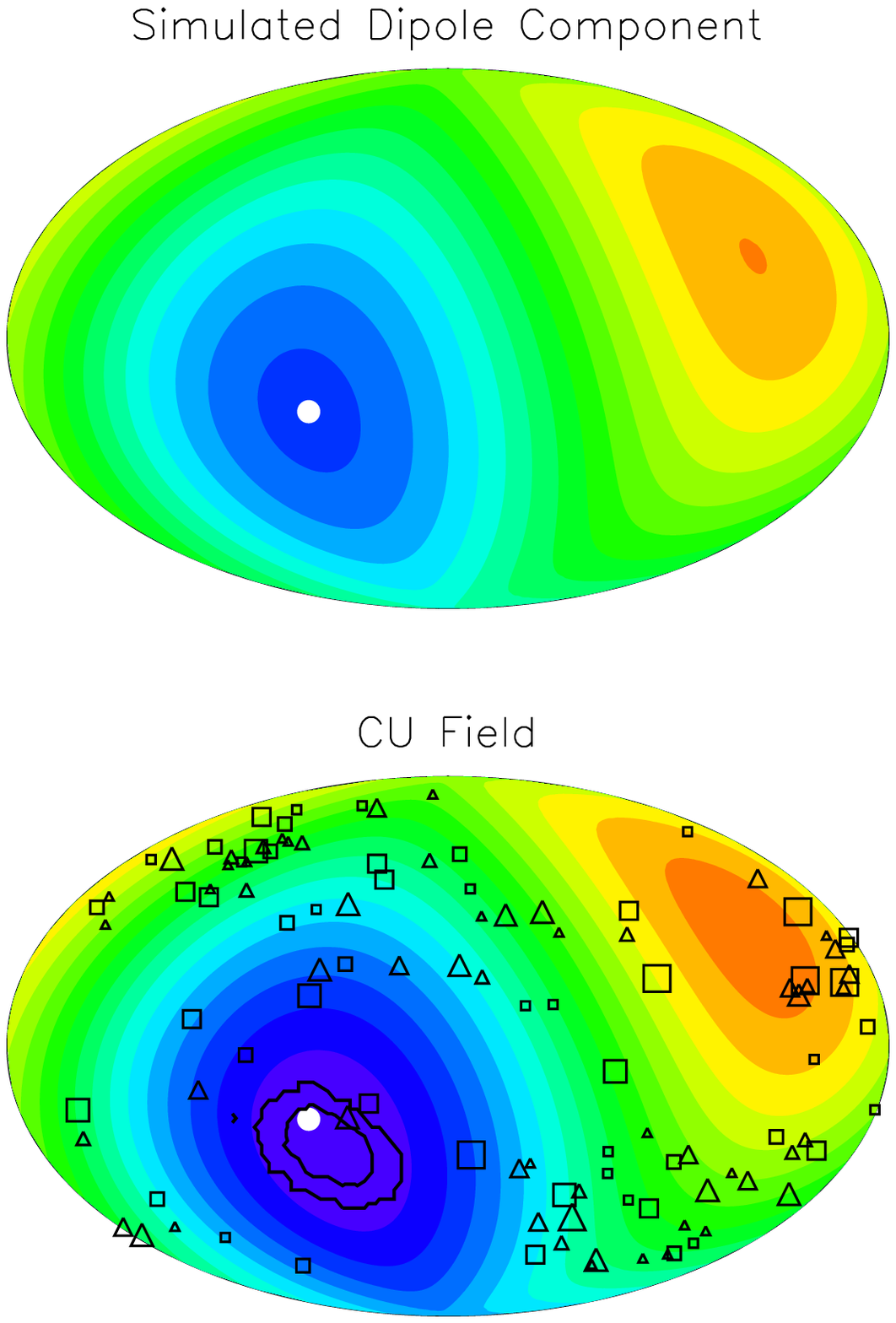}
\includegraphics[totalheight=2.5in]{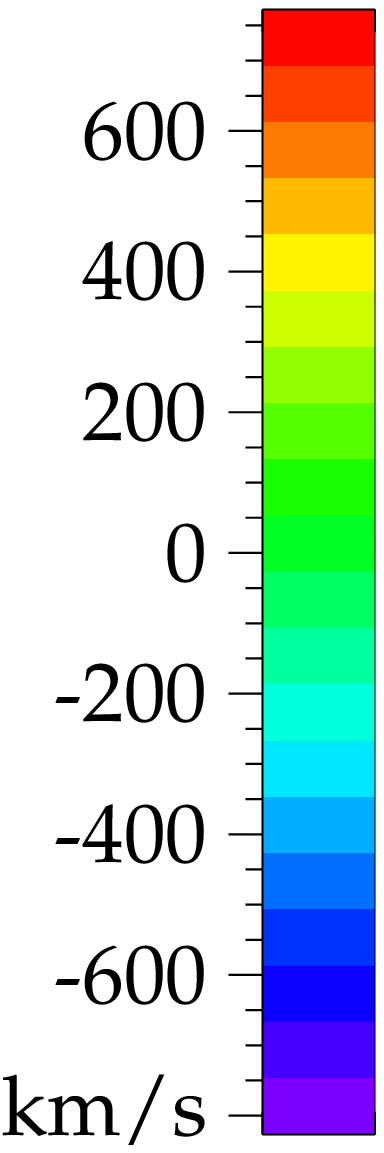}
\caption{Simulated velocity field for Case 1 (top left), dipole component of that field (top right), WLS dipole result (bottom left) and CU dipole result (bottom right) for a typical simulation of 200 data points.  Data points are overlaid as triangles (positive peculiar velocity) and squares (negative peculiar velocity).  Error contours (68\% and 95\%) are marked as black lines. The 95\% contour for CU and WLS enclose the direction of the true dipole, marked as a white circle.  The WLS result is more representative of the actual field while CU has a more accurate dipole.}
\label{fig:case1}
\end{center}
\end{figure}

For Case 1 the true dipole points along a sparsely sampled direction (Fig.~\ref{fig:case1}).  In the top row are the simulated velocity field and the dipole component of that field.  On the bottom are the results for CU and WLS.  In all plots the true direction of the dipole is shown as a white circle.  
 As WLS is optimized to model the velocity field, it is no surprise that WLS overestimates the magnitude of the dipole (bottom left) to better model the simulated velocity field (top left).  This behavior is very similar to what we saw in \S \ref{sec:sim}.  WLS is aliasing power onto different scales to best model the field, sacrificing unbiased coefficients.  
If we compare the CU velocity field to the simulated velocity field we see that it is less accurate but that CU's estimate of the dipole is a more accurate measure of the true dipole (top right).  Both methods are recovering the direction of the dipole at roughly the 2$\sigma$ level, leading us to conclude that it is the magnitude of the dipole which is most variable between the methods for this case.  
 
\begin{figure}[ht]
\begin{center}
\includegraphics[totalheight=7in]{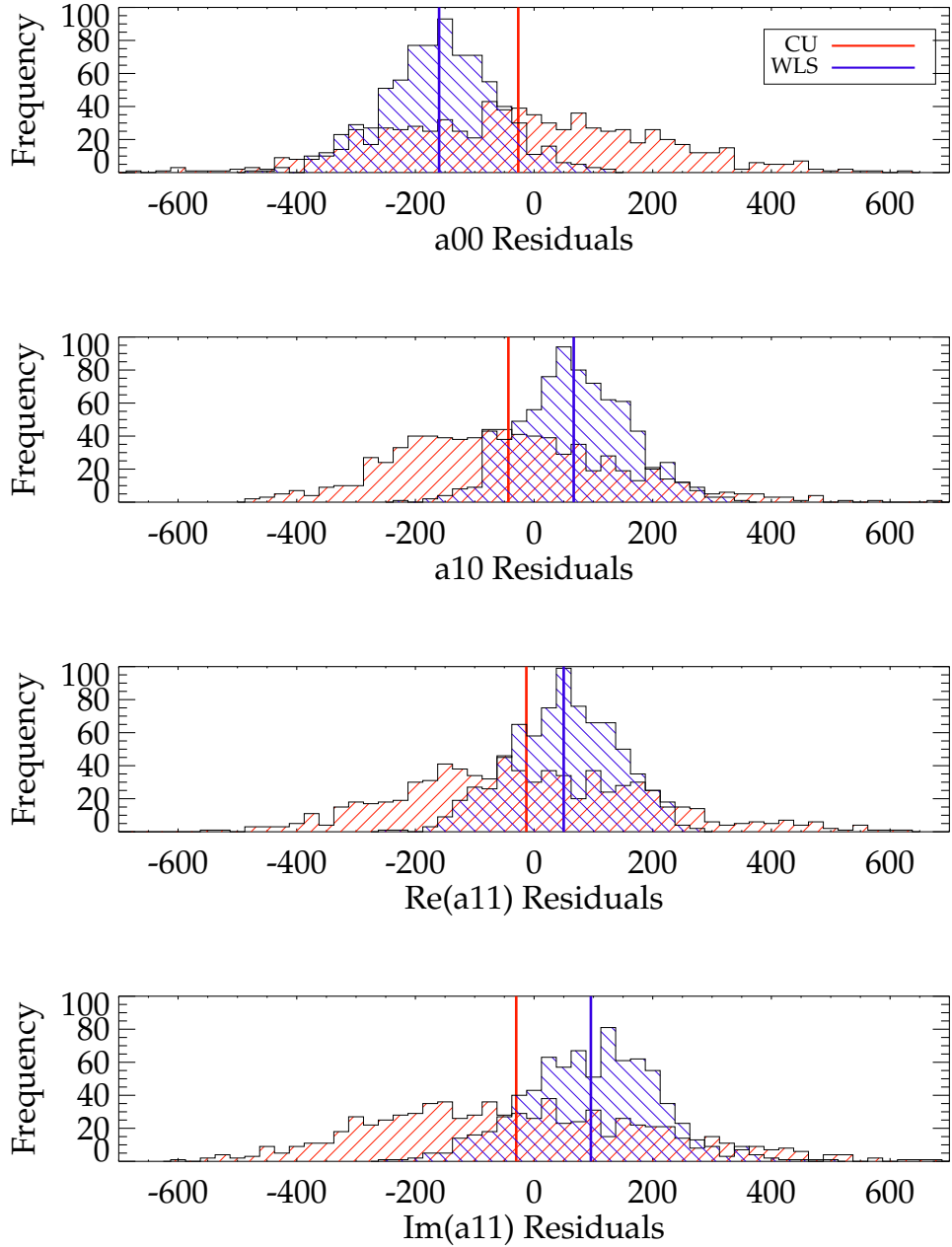}
\caption{Distribution of CU (WLS) coefficients minus the true values for Case 1 in red (blue) for 770 simulated datasets.  The vertical lines indicate the mean of the distribution.  The distance the mean is from zero is an indication of the bias.  The spread in the distributions indicates the uncertainty.  WLS is more biased than CU but CU has larger uncertainties.}
\label{fig:diffhist1}
\end{center}
\end{figure}

One may more easily see the difference in WLS and CU determined coefficients in Figure~\ref{fig:diffhist1}, where we plot
the difference distributions (regression determined coefficients minus the true coefficients) for 770 simulations of 200 data points.  
Ideally these distributions would be centered at zero with a narrow spread.  The distance the mean of the distribution is from zero is an indication of the bias.  The spread is an indication of the error. 
We see that WLS is more biased than CU but the uncertainty in CU is much larger.  

\begin{figure}[ht]
\begin{center}
\includegraphics[totalheight=3.5in]{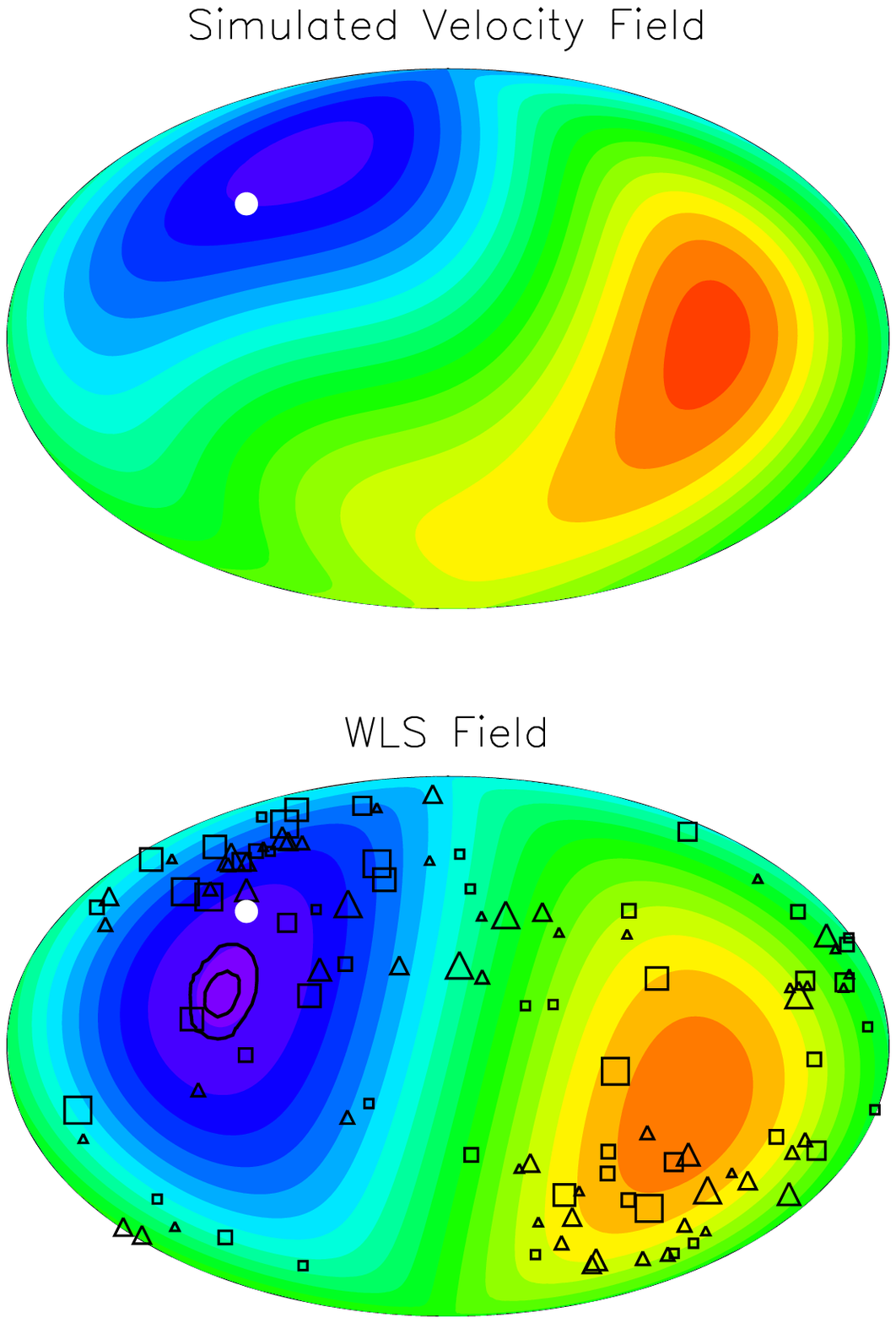}
\includegraphics[totalheight=3.5in]{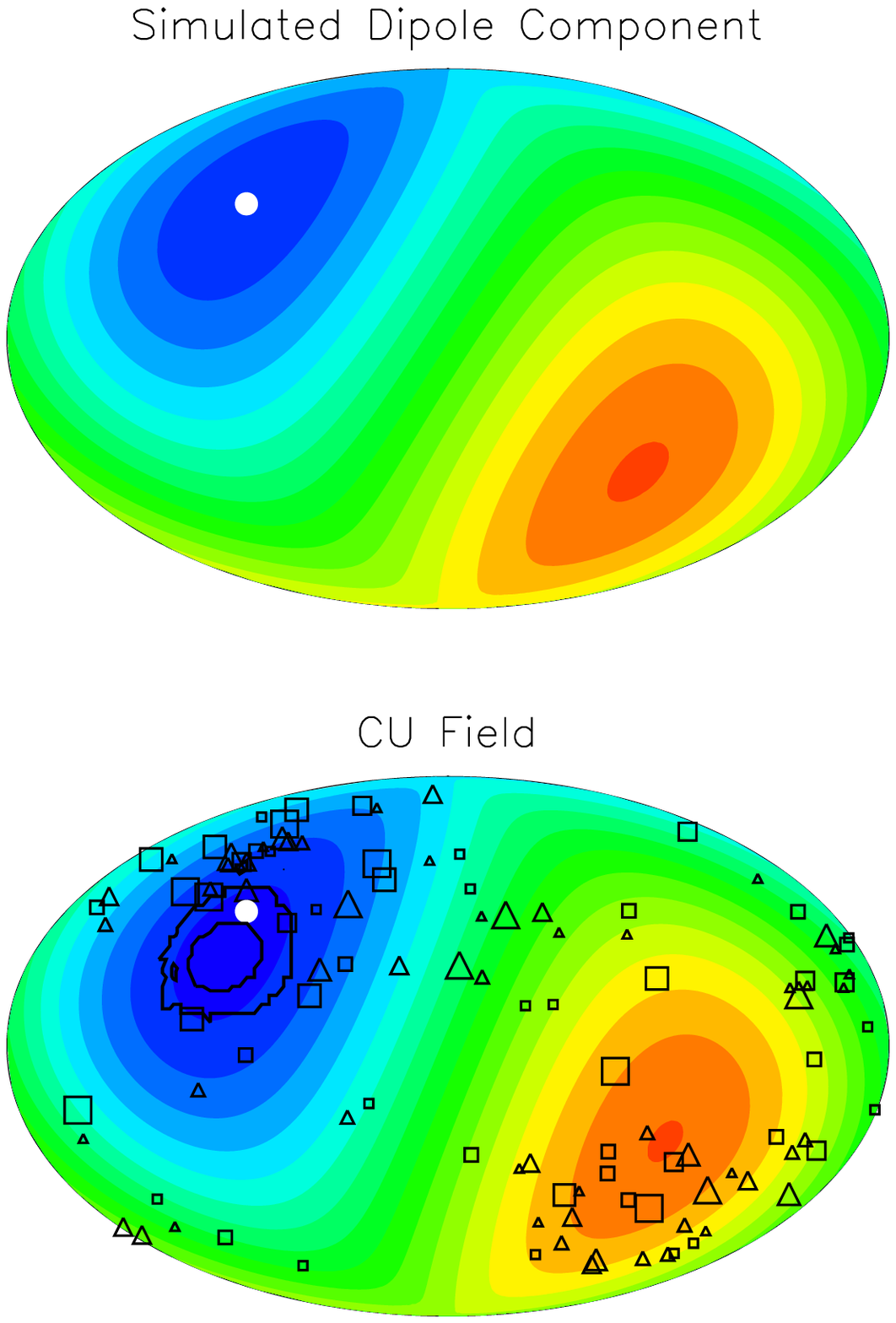}
\includegraphics[totalheight=2.5in]{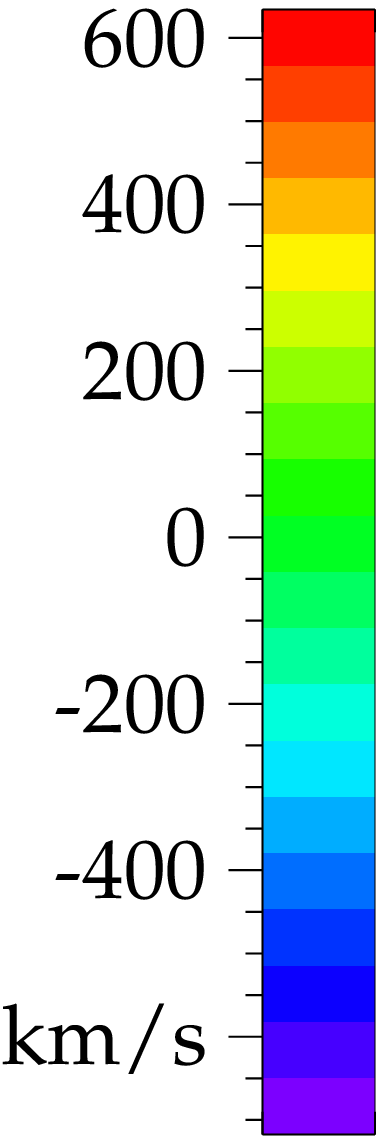}
\caption{Simulated velocity field for Case 2 (top left), dipole component of that field (top right), WLS dipole result (bottom left) and CU dipole result (bottom right) for a typical simulation.  Data points are overlaid as triangles (positive peculiar velocity) and squares (negative peculiar velocity).  In this scenario, the 95\% contour for WLS, marked in black, completely misses the direction of the true dipole, marked as the white circle.  The WLS dipole is pulled toward a region of space less sampled.}
\label{fig:case2}
\end{center}
\end{figure}

\begin{figure}[ht]
\begin{center}
\includegraphics[totalheight=7in]{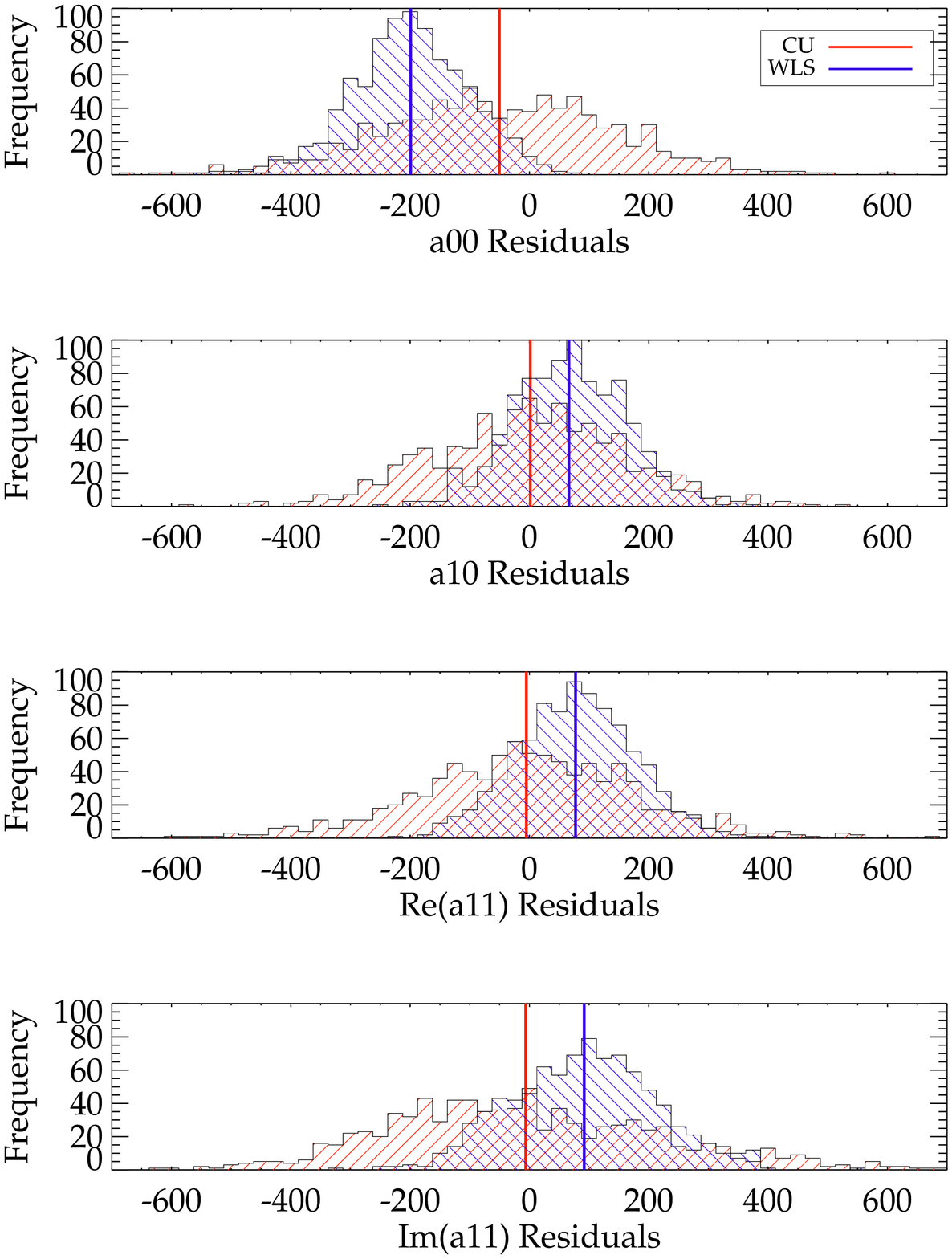}
\caption{Distribution of CU (WLS) coefficients minus the true values for Case 2 in red (blue) for 874 simulated datasets.  The vertical lines indicate the mean of the distribution. WLS is consistently more biased than CU}
\label{fig:diffhist2}
\end{center}
\end{figure}

In Case 2 the true dipole points along a region of space which is densely sampled (Fig.~\ref{fig:case2}).  In the bottom left plot we see the direction of the dipole for WLS is pulled down toward a region of space which is less sampled.  Since the true direction of the dipole is well constrained by data, to more accurately model the flow field WLS must alter the direction of the dipole toward a less sampled region.  This is necessary as WLS is trying to account for power which is really part of the quadrupole with the dipole term.  As a result, WLS misses the true direction of the dipole at the 95\% confidence level.   This may be similar to what we see in Fig.~\ref{fig:contourWLSCU} where the WLS dipole points more along the galactic plane when compared to CU.  CU is less sensitive to this affect as it is optimized to find unbiased coefficients.  Correspondingly, CU encloses the true direction of the dipole at the 95\% confidence level.  In Figure~\ref{fig:diffhist2} we plot the difference distributions as we did for Case 1.  It is clear that the WLS coefficients are more biased than CU but that the uncertainty in CU is much larger.  

We can explicitly check the bias of the methods using the simulated data of \S \ref{sec:dipoleAnalysis}.  The important calculation is the probability that the 95\% confidence interval for a given simulation includes the true value.  For an accurately determined confidence interval, this should happen 95\% of the time.  We start with one simulated dataset and perform 1000 bootstrap resamples.  This gives us distributions of the coefficients from which we can determine the confidence intervals.  We then determine if the true values falls within this interval.  After doing this for all of the simulations from \S \ref{sec:dipoleAnalysis}, we can measure how often the true value falls within the confidence interval.  These probabilities are summarized in Table \ref{table:confInt}.  

CU is more accurate in its estimate of the 95\% confidence interval for both cases.  
The lower probabilities for WLS are a result of the bias in the method.  By construction, the WLS confidence intervals are centered about the regression-determined coefficients.  If the coefficients are biased, the WLS intervals are shifted and the true value will lie outside this interval more often than expected. 

\begin{center}
\begin{deluxetable}{ccccc}
\tablewidth{0pt}
\tablecolumns{5}
\tablecaption{Probability of the 95\% confidence interval containing the truth}
\tablehead{\multicolumn{1}{c}{} & \multicolumn{1}{c}{$a_{00}$} & \multicolumn{1}{c}{$a_{10}$} & \multicolumn{1}{c}{$\Re (a_{11})$} & \multicolumn{1}{c}{$\Im (a_{11})$}}
\startdata
Case 1 & & & &\\
WLS & 0.50   &   0.88   &   0.86 &     0.88 \\
CU & 0.93   &   0.90   &   0.89   &   0.92 \\
Case 2 & & & & \\
WLS   &   0.49  &   0.88   &  0.86   &   0.88\\
CU    &   0.93  &   0.94  &   0.94   &   0.91\\
\enddata
\label{table:confInt}
\end{deluxetable}
\end{center}

\section{Conclusion}\label{sec:conclusion}

In this work, we applied statistically rigorous methods of nonparametric risk estimation to the problem of inferring the local peculiar velocity field from nearby SNIa.  We use two nonparametric methods - WLS and CU - both of which employ spherical harmonics to model the field and use the risk to determine at which multipole to truncate the series.  The minimum of the estimated risk will tell one the maximum multipole to use in order to achieve the best combination of variance and bias.  The risk also conveys which method models the data most accurately.

WLS estimates the coefficients of the spherical harmonics via weighted least squares.  We show that if the data are not drawn from a uniform distribution and if there is power beyond the maximum multipole in the regression, WLS fitting introduces a bias on the coefficients.  CU estimates the coefficients without this bias, thereby modeling the field over the entire sky more realistically but sacrificing in accuracy. Therefore, if one believes there is power beyond the tuning parameter or the data are not uniform, CU may be more appropriate when estimating the dipole, but WLS may describe the data more accurately.

After applying nonparametric risk estimation to SNIa we find that there is not enough data at this time to measure power beyond the dipole.  There is also no significant evidence of a monopole term for either WLS or CU, indicating that we are using consistent values of $H_0$ and $M_V$.  The WLS Local Group bulk flow is moving at $538 \pm 86$~km~s$^{-1}$ towards ($l,b$) $=$ ($258^{\circ} \pm 10^{\circ}, 36^{\circ} \pm 11^{\circ}$) and the CU bulk flow is moving at $446 \pm 101$~km~s$^{-1}$ towards ($l,b$) $=$ ($273^{\circ} \pm 11^{\circ}, 46^{\circ} \pm 8^{\circ}$).  After performing a paired t-test we find that these values are in agreement.

To test how CU and WLS perform on a more realistic dataset, we simulate data similar to the actual data and investigate how they perform as we change the direction of the dipole.  We find for our two test cases, that CU produces less biased coefficients than WLS but that the uncertainties are larger for CU.  We also find that the 95\% confidence intervals detemined by CU are more representative of the actual 95\% confidence intervals.  

We estimate using simulations that with $\sim$200 data points, roughly double the current sample, we would be able to measure the quadrupole moment assuming a similarly distributed dataset. Nearby SNIa programs such as the CfA Supernova Group, Carnegie Supernova Project, KAIT, and the Nearby SN Factory will easily achieve this sample size in the next one to two years.   The best way to constrain higher-order moments however, would be to obtain a nearly uniform distribution of data points on the sky.  \citet{Haugbolle07} estimate that with a uniform sample of 95 SNIa we can probe $l=3$ robustly.  

With future amounts of data the analysis can be expanded not only out to higher multipoles, but to modeling the peculiar velocity field as a function of redshift.  This will enable us to determine the redshift at which the bulk flow converges to the rest frame of the CMB.  Binning the data in redshift will also allow one to look for a monopole term that would indicate a Hubble bubble. 

As there is no physical motivation for using spherical harmonics to model the sampling density, future increased amounts of data will also allow us to use nonparametric kernel smoothing both to estimate $h$ and the peculiar velocity field; this would be ideal for distributions on the sky which subtend a small angle, like the SDSS-II Supernova Survey sample \citep{Sako08, Frieman08}.  

As astronomical data sets continue to grow, it becomes increasingly important to pursue statistical methods like those outlined in this work.

\acknowledgments
We would like to thank Malcolm Hicken for sharing the consistent results and RPK for his detailed useful comments.
We would also like to thank Jeff Newman for useful discussions and the referee for useful suggestions.
This research has made use of the NASA/IPAC Extragalactic Database (NED) which is operated by the Jet Propulsion Laboratory, California Institute of Technology, under contract with the National Aeronautics and Space Administration.

\appendix
\section{Appendix}

\subsection{Bias on WLS coefficients $\hat \beta_J$}\label{appendix:biasWLS}
To determine the bias on the estimated coefficients, $\hat \beta_J$, recall that we can model any velocity field with an infinite set of spherical harmonics
\begin{equation}\label{U}U=Y\beta+\epsilon\end{equation}
where $\beta$ is the column vector given by $\beta = (\beta_0...\beta_{\infty})$ and 
\begin{equation}
Y=
\left[ {\begin{array}{cccc}
\phi_0(x_1) & \phi_1(x_1) & \cdots&\phi_\infty(x_1)\\
\phi_0(x_2)&\phi_1(x_2)&\cdots&\phi_\infty(x_2)\\
\vdots&\vdots &\cdots &\cdots\\
\phi_0(x_N)&\phi_1(x_N)&\cdots &\phi_\infty(x_N)\\
\end{array} } \right]
\end{equation}
If we substitute $U$ into Eq.~\ref{betahat} we get
\begin{equation}\hat\beta_J=(Y_J^TWY_J)^{-1}\,Y_J^T\,W\,(Y\beta+\epsilon)\end{equation}
If we decompose $Y\beta$ into
\begin{equation}Y\beta=Y_J\beta_J+Y_{\infty}\beta_{\infty}\end{equation}
where $\beta_{\infty} = (\beta_{J+1}...\beta_{\infty})^T$ and 
\begin{equation}
Y_{\infty}=
\left[ {\begin{array}{cccc}
\phi_{J+1}(x_1) & \phi_{J+2}(x_1) & \cdots&\phi_{\infty}(x_1)\\
\phi_{J+1}(x_2)&\phi_{J+2}(x_2)&\cdots&\phi_\infty(x_2)\\
\vdots&\vdots &\cdots &\cdots\\
\phi_{J+1}(x_N)&\phi_{J+2}(x_N)&\cdots &\phi_\infty(x_N)\\
\end{array} } \right]
\end{equation}
 then 
\begin{eqnarray}
\hat\beta_J&=&(Y_J^TWY_J)^{-1}\,Y_J^T\,W\,(Y_J\beta_J+Y_{\infty}\beta_{\infty}+\epsilon)\\
&=&\beta_J+(Y_J^TWY_J^T)^{-1}\,Y_J^T\,W\,(Y_{\infty}\beta_{\infty}+\epsilon)
\end{eqnarray}
The bias on $\hat\beta_J$ is
\begin{eqnarray}
\beta_J-\left\langle \hat\beta_J \right\rangle &=& \beta_J - \left\langle \beta_J+(Y_J^TWY_J)^{-1}\,Y_J^T\,W\,(Y_{\infty}\beta_{\infty}+\epsilon) \right\rangle\\
&=& \left\langle (Y^TWY)^{-1}\,Y^T\,W\,(Y_{\infty}\beta_{\infty}) \right\rangle .
\end{eqnarray}

\subsection{Bias on CU Coefficients $\hat \beta^*$}\label{appendix:biasCU}

To determine the bias for the weighted coefficients $\hat\beta^*$ we first multiply the top and bottom of Eq.~\ref{CUbeta} by $1/N$
\begin{eqnarray}
\hat\beta_j^* &=& \frac{\frac{1}{N}\displaystyle\sum_{n=1}^N{\frac{U_n\phi_j(x_n)}{h(x_n)\sigma_n^2}}}{\frac{1}{N}\displaystyle\sum_{n=1}^N \frac{1}{\sigma_n^2}}  \\
&=& \frac{ \left \langle \frac{U_n\phi_j(x_n)}{h(x_n)\sigma_n^2} \right \rangle}{\left \langle\frac{1}{\sigma_n^2} \right \rangle}
\end{eqnarray}
If $U(x_n)$, $\phi_j(x_n)$, and $h(x_n)$ are all independent of $\sigma_n$, then
\begin{eqnarray}
\left \langle \hat \beta_j^* \right \rangle &=& \frac{\left \langle \frac{U_n\phi_j(x_n)}{h(x_n)\sigma_n^2} \right \rangle}{\left \langle \frac{1}{\sigma_n^2} \right \rangle}\\
&=&\frac{\left \langle \frac{U_n\phi_j(x_n)}{h(x_n)} \right \rangle \left \langle\frac{1}{\sigma_n^2} \right \rangle}{\left \langle \frac{1}{\sigma_n^2} \right \rangle}\\
&=& \left \langle \frac {U(x_n)\phi_j(x_n)}{h(x_n)} \right \rangle\\
&=&\beta_j
\end{eqnarray}
So our bias is $\beta - \langle \beta_j^* \rangle = 0$. 


\subsection{Risk Estimation}\label{appendix:risk}

The risk is a way of determining how many basis functions should be in $f(x)$ and can be written as 
\begin{equation}\label{R}R= \left\langle (\hat\theta-\theta)^2 \right\rangle \end{equation}
where $\hat\theta$ is the estimated or measured value of some true parameter, $\theta$.  The expectation value of $\hat\theta$ is the mean, $\bar \theta$
\begin{equation} \bar\theta  \equiv  \langle \hat\theta \rangle  \end{equation}
By adding and subtracting the mean from $(\hat\theta-\theta)$ in Eq.~\ref{R}, the risk can be written in terms of the variance and the bias.
\begin{eqnarray}
R&=& \left\langle (\hat\theta-\bar\theta+\bar\theta-\theta)^2 \right\rangle\\
&=& \left\langle (\hat\theta-\bar\theta)^2 \right\rangle+(\bar\theta-\theta)^2+(\bar\theta-\theta) \left\langle (\hat\theta-\bar\theta) \right\rangle\\
&=&\left\langle (\hat\theta-\bar\theta)^2 \right\rangle +(\bar\theta-\theta)^2\\
&=& \rm {Var}(\hat\theta)+ \rm {bias}^2
\end{eqnarray}

\subsection{Smoothing Matrix for CU Regression}\label{appendix:L}

\begin{eqnarray}
\hat f(x_i) &=& \displaystyle\sum_{j=0}^J\hat\beta_j^*\phi_j(x_i)\\
&=&\displaystyle\sum_{j=0}^J\phi_j(x_i) \frac{\displaystyle\sum_{n=1}^N\frac{U_n\phi_j(x_n)}{h(x_n)\sigma_n^2}}{\displaystyle\sum_{n=1}^N \frac{1}{\sigma_n^2}}\\
&=& \displaystyle\sum_{n=1}^N U_n \frac{\displaystyle\sum_{j=0}^J\frac{\phi_j(x_i)\phi_j(x_n)}{h(x_n)\sigma_n^2}}{\displaystyle\sum_{n=1}^N \frac{1}{\sigma_n^2}} \\
&=& LU
\end{eqnarray}

\bibliographystyle{astroads}
\bibliography{bib}


\end{document}